\documentclass[useAMS,usenatbib]{mn2e}
\usepackage[dvips]{graphicx}
\usepackage{times}
\usepackage{color}
\usepackage{wrapfig}
\usepackage{float}
\usepackage{array}
\usepackage{subcaption}
\usepackage{graphicx}

\long\def\symbolfootnote[#1]#2{\begingroup%
\def\thefootnote{\fnsymbol{footnote}}\footnote[#1]{#2}\endgroup}
\renewcommand{\thefootnote}{\fnsymbol{footnote}}

\begin{document}
 \vspace{-5cm}
\title[CO observations of local galaxy pairs]
{Galaxy pairs in the SDSS - XIII.  The connection between enhanced star formation and molecular gas properties in galaxy mergers. }

\author[Violino et al.] 
{\parbox[h]{\textwidth}{
Giulio~Violino,$^{1}$   
Sara~L.~Ellison,$^{2}$
Mark~Sargent,$^{3}$ 
Kristen~E.~K.~Coppin,$^{1}$\\
Jillian~M.~Scudder,$^{4}$
Trevor~J.~Mendel $^{5}$ \&
Amelie~Saintonge$^{6}$
}
\vspace*{6pt}\\ 
$^{1}$ Centre for Astrophysics Research, University of Hertfordshire, College Lane, Hatfield, AL10 9AB, UK. \\
$^{2}$ Department of Physics $\&$ Astronomy, University of Victoria, Finnerty Road, Victoria, British Columbia, V8P 1A1, Canada \\
$^{3}$  Astronomy Centre, Department of Physics and Astronomy, University of Sussex, Brighton, BN1 9QH, UK \\
$^{4}$ Department of Physics and Astronomy, Oberlin College, Oberlin, Ohio, 44074, USA \\ 
$^{5}$ Max-Planck-Institute fur Extraterrestrische Physik, Giessenbachstrasse, D-85748 Garching, Germany \\
$^{6}$  Department of Physics and Astronomy, University College London, Gower Street, London, WC1E 6BT, UK\\
} 

\maketitle

\begin{abstract}
We investigate the connection between star formation and molecular gas
properties in galaxy mergers at low redshift (z$\leq$0.06). The study
we present is based on IRAM 30-m CO(1--0) observations of 11 galaxies
with a close companion selected from the Sloan Digital Sky Survey
(SDSS).  The pairs have mass ratios $\leq$4, projected separations
r$_{\mathrm{p}} \leq$30 kpc and velocity separations
$\Delta$V$\leq$300 km s$^{-1}$, and have been selected to exhibit
enhanced specific star formation rates (sSFR).  We calculate molecular
gas (H$_{2}$) masses, assigning to each galaxy a physically motivated
conversion factor $\alpha_{\mathrm{CO}}$, and we derive molecular gas
fractions and depletion times. We compare these quantities with those
of isolated galaxies from the extended CO Legacy Data base for the
GALEX Arecibo SDSS Survey sample (xCOLDGASS, \citealt{Saintonge17})
with gas quantities computed in an identical way.  Ours is the first
study which directly compares the gas properties of galaxy pairs and
those of a control sample of normal galaxies with rigorous control
procedures and for which SFR and H$_{2}$ masses have been estimated
using the same method.  We find that the galaxy pairs have shorter
depletion times and an average molecular gas fraction enhancement of
0.4 dex compared to the mass matched control sample drawn from
xCOLDGASS.  However, the gas masses (and fractions) in galaxy pairs
and their depletion times are consistent with those of non-mergers
whose SFRs are similarly elevated. We conclude that both external
interactions and internal processes may lead to molecular gas
enhancement and decreased depletion times.

\end{abstract}
\begin{keywords}
galaxies: ISM, galaxies: interactions, galaxies: evolution, radio lines: galaxies  

\end{keywords}

\section{Introduction}

Galaxy interactions represent a fundamental component of our current view of hierarchical galaxy evolution. 
Studies based on both observations and simulations have shown that galaxy collisions and mergers can dramatically affect the galaxies undergoing the interaction, by e.g., triggering nuclear activity (e.g. \citealt{Kennicutt84}; \citealt{Kennicutt87}; \citealt{Ellison11}, 2013a; \citealt{Silverman11}, \citealt{Satyapal14}), producing colour changes (e.g. \citealt{Larson78}; \citealt{Darg10}; \citealt{Patton11}), disrupting morphologies (e.g. \citealt{Kaviraj11}; \citealt{Patton16}  \citealt{Lofthouse17}) and altering the metallicities (e.g. \citealt{Rupke10}; \citealt{Perez11}; \citealt{Scudder12}; \citealt{Torrey12}).
The most evident effect driven by galaxy encounters is probably the triggering of new episodes of star formation, which can occur both in the pre-merger regime between first pericentre and coalescence (e.g. \citealt{Nikolic04}; \citealt{Patton11}; \citealt{Scudder12}; Ellison et al., 2008a, 2013b), and in the post-merger phase, where the two nuclei of the interacting galaxies have merged together (e.g. \citealt{Ellison13a}; Kaviraj et al. 2012, 2014).   
The idea that galaxy mergers have a strong impact on the star formation activity is supported by studies of Ultra-Luminous InfraRed Galaxies (ULIRGs), i.e. galaxies with IR luminosities exceeding 10$^{12}$ L$_{\odot}$ and characterized by SFRs up to $\sim$1000 M$_{\odot}$yr$^{-1}$ (e.g. \citealt{Mihos94}; \citealt{Barnes91}; \citealt{Daddi10}; \citealt{Scoville15}). Observations have revealed that the majority of ULIRGs reside in interacting systems (e.g. \citealt{Sanders96}; \citealt{Veilleux02}; Kartaltepe et al. 2010, 2012; \citealt{Haan11}). 
 Nevertheless, ULIRGs are rare and extreme examples of highly star-forming galaxies. Most galaxy-galaxy interactions result in SFR increases of at most a factor of a few, as shown in both numerical simulations (e.g. \citealt{DiMatteo08}) and observations of galaxy pairs and post-mergers (\citealt{Ellison08}; \citealt{Martig08}; \citealt{Jogee09}; \citealt{Robaina09}; \citealt{Scudder12}).
 
Theoretical work on galaxy encounters suggests that there are two main factors responsible for the enhancement of the star formation during a merger event. The first is an enrichment of the molecular gas reservoir available for fuelling  star formation. This increase in the H$_{2}$ fraction can be explained by invoking an accelerated transition from atomic (HI)
to molecular gas due to collision-induced external pressure (\citealt{Kaneko13b}).
\cite{Moster11} presents a physically motivated scenario for explaining this phenomenon from
 a set of cosmological hydrodynamical simulations of major mergers for which they include both a gas disk and a gas halo. This last component firstly drifts towards the centre of the galaxy and consequently cools down, causing a growth of the H$_{2}$ content.
The second driver of enhanced star formation in mergers 
is an increase of the density of molecular gas, which induces a more efficient conversion of gas into stars.
Indeed, numerical and hydrodynamical simulations predict that during the merger, gravitational torque decreases the angular momentum of gas which flows towards the galactic centre; the result is a rapid increase of the gas density which finally brings about a burst of nuclear star formation (\citealt{Mihos96}; \citealt{DiMatteo08}; \citealt{Renaud14}).
Besides nuclear starbursts, interactions can also trigger
highly efficient star formation across the whole galaxy through several episodes of gas fragmentation in dense clouds induced by gravitational torques, as was shown in high-resolution hydrodynamical simulations (\citealt{Teyssier10}, \citealt{Bournaud11}).

In order to test these theoretical predictions, numerous observational studies have investigated 
how the molecular gas content and the  star formation activity are influenced by galaxy interactions, and how these vary across different phases of the merger.
However, the majority of these have been hampered by several factors, including limited statistics (e.g. \citealt{Braine04} and \citealt{Boquien11} only studied single galaxy interactions), heterogeneous samples (often being a mix of pre-mergers and merger remnants, e.g. \citealt{Braine93}, \citealt{Casasola04}), and a lack of suitable control samples (e.g. \citealt{Michiyama16} used a comparison sample of only few sources with measurements of CO(3--2) from \citealt{Tacconi13}).
% The first is a limited statistics. That is the case, e.g., of 
%Secondly, the samples which have been considered were often not homogeneous, being a mix of both galaxies in a pre-merger phase 
%and merger-remnants (e.g. ).
%In addition, most of the previous studies lacked a suitable comparison sample large enough to allow a robust
%assessment of galaxy pairs gas properties. For instance,  analysed the gas properties and star-formation activity of a sample of early- and late-state %mergers; however, their study is based on CO(3--2) transitions which makes their data only directly comparable with few previous measurements of CO(3--2) %(e.g. \citealt{Tacconi13}).

Another complicating factor is the lack of a physically motivated conversion factor $\alpha_{\mathrm{CO}}$ between the measured CO luminosity and molecular gas mass. In fact, a standard disk-like value 
has often been used for mergers
($\alpha_{\mathrm{CO}}$=3.2 e.g. \citealt{Combes94}), which may not be appropriate, given that the interaction is capable of altering the ISM condition and morphology in the merging galaxies, which could result in a different relation between CO emission and H$_{\mathrm{2}}$ content.
In addition, more recent studies show that the conversion factor is not universal, varying from one source to another by up to a factor of $\sim$10$^{3}$, depending on the gas surface density, metallicity and stellar mass (e.g. \citealt{Narayanan12}; \citealt{Bolatto13}).

In this paper, we tackle these previous shortcomings by carefully selecting a sample of only galaxy pairs, adopting a physically motivated CO--H$_{\mathrm{2}}$ conversion factor and making use of the new extended
GALEX Arecibo SDSS Survey (xCOLDGASS; \cite{Saintonge17}) to build a suitable comparison sample.
We have carried out a systematic study of the molecular gas content (H$_{\mathrm{2}}$), as traced by both the CO(1--0) and CO(2--1) using the IRAM 30-m,
and the star formation activity
of 11 galaxy pairs.
Our main goal is to investigate the effects of galaxy interactions on the molecular gas component of the galaxies undergoing
a merger, prior to the final coalescence stage (we study the molecular gas content of post-merger galaxies in a forthcoming paper: Sargent et al., in prep). Specifically, we want to test whether the star formation enhancement exhibited by the galaxy pairs is related to either an enrichment
of the gas content or to a  decrease of the gas depletion time, or both.  

This paper is organized as follows: in Section 2 we describe the sample selection of our galaxy pairs,
while in Section 3 we describe the IRAM 30-m CO observations and data reduction. In Section 4 we present the analysis and our main results, which we discuss in Section 5. 
Finally, our conclusions are presented in Section 6 together with proposed future work to expand this study. 
Throughout this paper we assume a Chabrier IMF and a flat $\Lambda$CDM cosmology with H$_{\mathrm{0}}$=69.6 km s$^{-1}$ Mpc$^{-1}$ and $\Omega_{\mathrm{M}}$=0.286 (\citealt{Wright06}).

\begin{figure}
\centering
\includegraphics[width=80mm,angle=0]{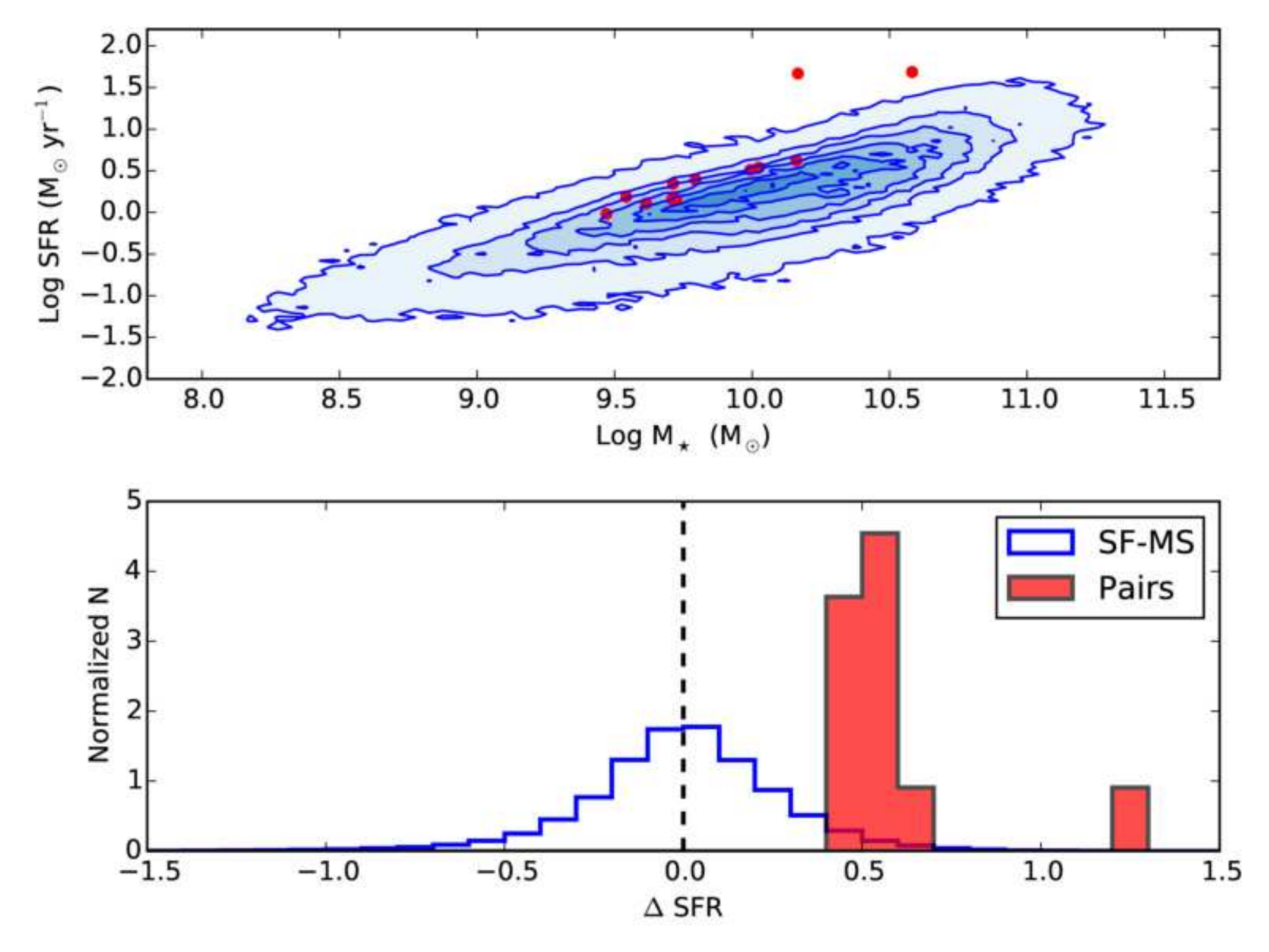}

\caption{Top panel: Comparison between the main-sequence of SDSS
  star-forming galaxies (blue contours, as classified by
  \citealt{Kauffmann03}) and our sample of 11 galaxies in pairs
  (filled red circles). Stellar masses and total SFRs of both samples
  are taken from Mendel et al. (2014) and \citet{Brinchmann04},
  respectively.  Bottom panel: Distribution of the SFR offset of
  galaxies in pairs compared to the main sequence, which reveals
  enhanced star formation activity in our sample of 11 galaxy
  pairs. The procedure used to produce this plot is fully described in
  \citet{Ellison16} and is summarized in Section 2.}
\end{figure}

\begin{figure*}
\centering
\includegraphics[width=1.2\textwidth]{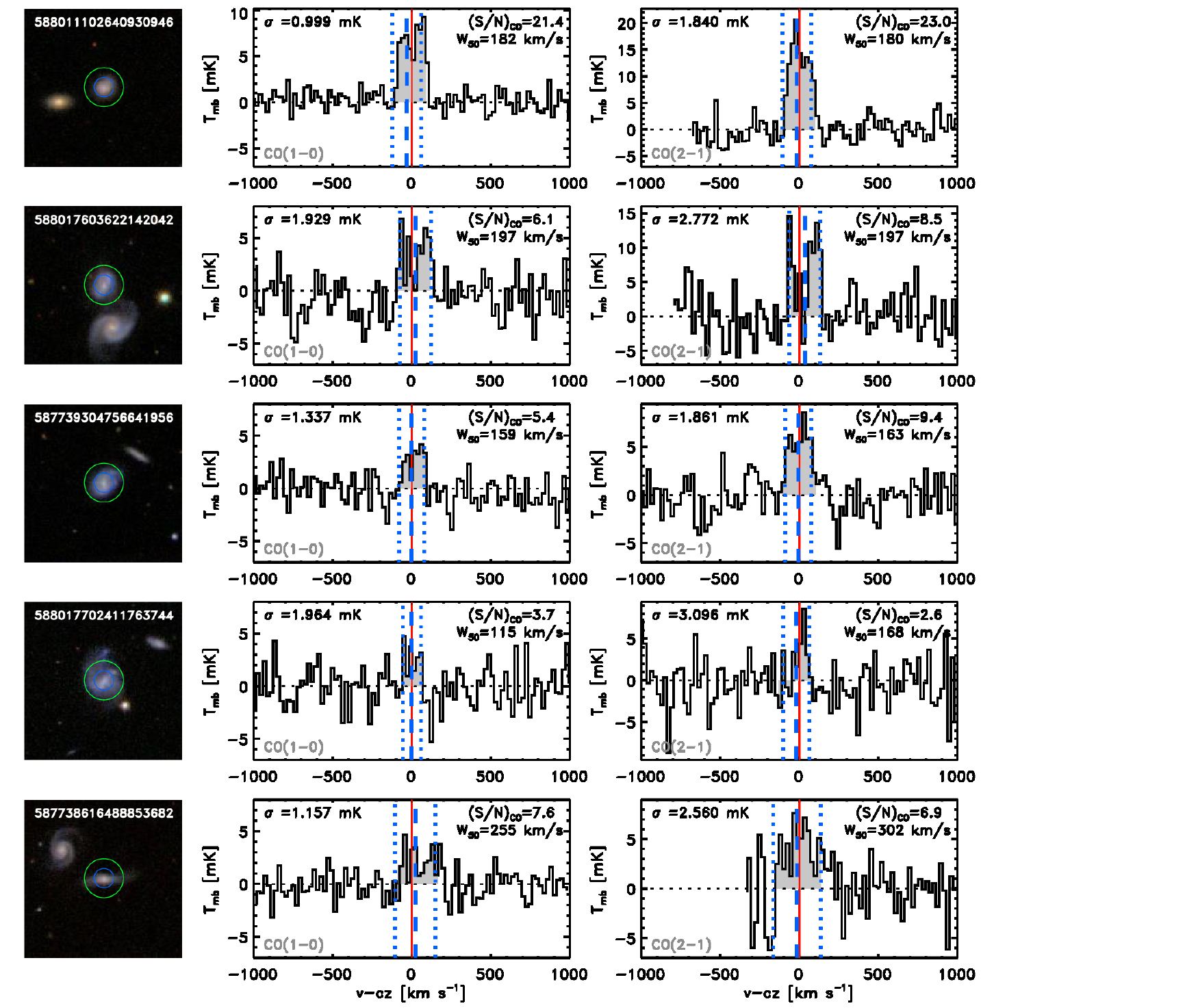}
\caption{SDSS cutout images of the 5 galaxy pairs in our sample and their corresponding CO(1--0) (left) and CO(2--1) spectra (right). The green and blue circles in the images represent the FWHM size of the IRAM 30-m beam of 22 and 11 arcsec, respectively. In each spectrum the red dashed line represents the systemic redshift of the source, as determined from the SDSS spectrum. The dashed blue line is the central velocity of the CO line, while the blue dotted line delimits  W50$_{\mathrm{CO}}$, i.e. the linewidth of the CO emission measured at half intensity.}
\end{figure*}

\begin{figure*}
\centering
\includegraphics[width=1.2\textwidth]{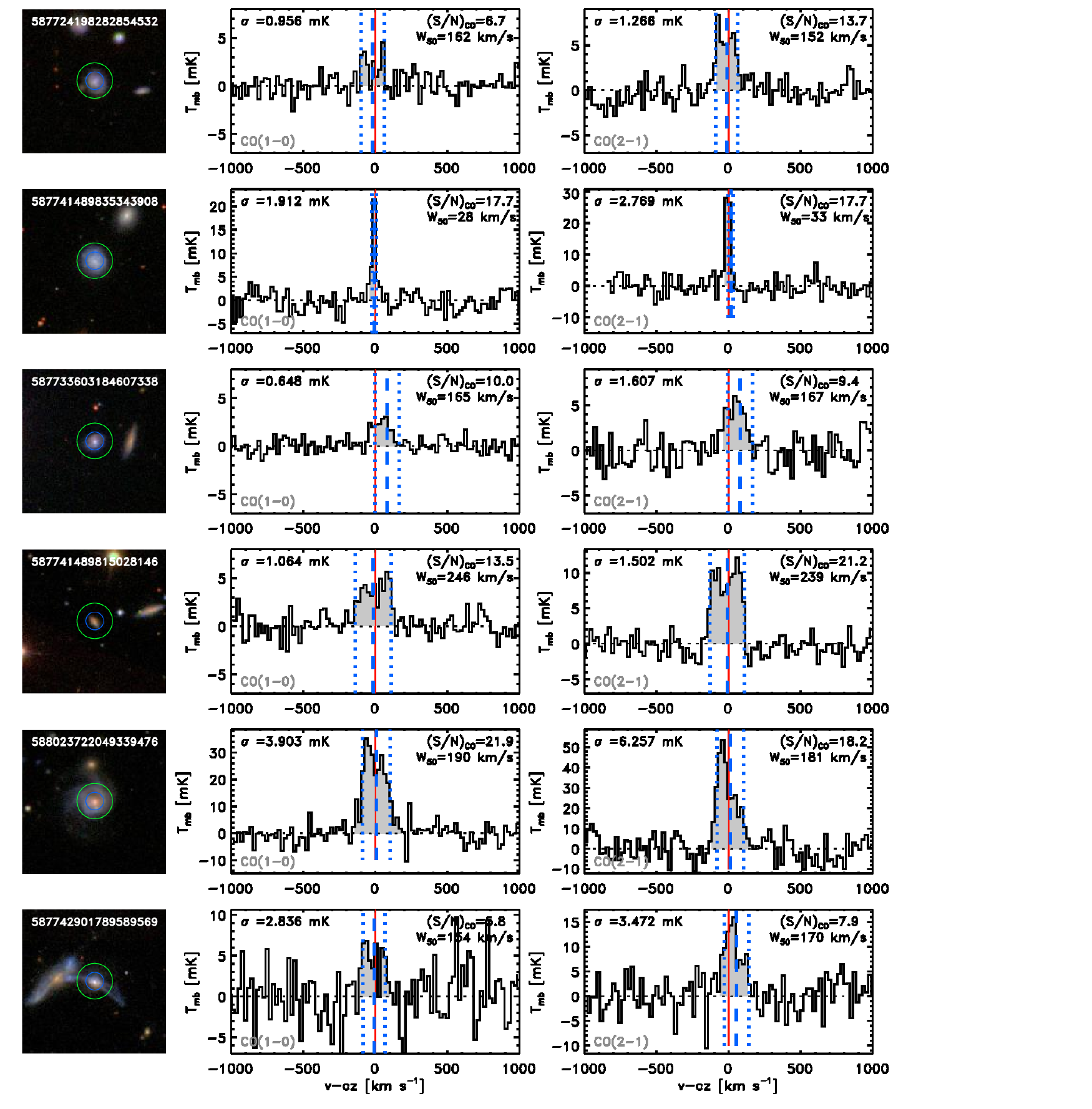}
\caption{SDSS cutout images of the remaining 6 galaxy pairs in our sample and their corresponding CO(1--0) (left) and CO(2--1) spectra (right). Details as in Figure 2.}
\end{figure*}

\begin{table*}
%\centering
\hspace{-0.2in}
%\begin{tabular}{K{5cm}K{5cm}K{2cm}K{2cm}K{2cm}K{2cm}K{2cm}K{2cm}K{2cm}}
\begin{tabular}{ccccccccc}
\hline
Source & SDSS DR7 objID &  M$_{*}$ & r$_{\mathrm{p}}$& $\Delta$V & mass ratio & redshift & logSFR & logSFR$^{\mathrm{aperture}}$  \\
&& [log M$_{\odot}$] & [kpc] &  [km s$^{-1}$] & & & [M$_{\odot}$yr$^{-1}$]& [M$_{\odot}$yr$^{-1}$]   \\
\hline
SDSSJ014845.47+134300.2 & 587724198282854532 &9.71& 26.59 &127.0 &2.57 & 0.045  &0.34   &  0.25  \\
SDSSJ080555.42+135959.0&587741489815028146&   9.79&26.78 &91.0& 0.61&0.038     & 0.40 & 0.39        \\
SDSSJ111633.85+284606.4& 587741489835343908& 9.47&16.66& 27.0 &0.72 & 0.024     &-0.02  & -0.11\\
SDSSJ112036.59+361234.4&587738616488853682 &9.99&29.71& 17.0& 0.73& 0.052	&0.51	&	0.46\\
SDSSJ123935.85+163516.1&587742901789589569& 9.72&19.55 &118.0& 3.90&0.026     &0.16 &    0.09  \\
SDSSJ125053.09+352404.9&587739304756641956 &9.62&16.55 &8.0& 2.69&0.033	& 0.11 &		0.03	\\
SDSSJ143154.09+215618.3&588023722049339476 &10.58&20.55 &13.0& 3.70&0.044     & 1.69 & 1.68        \\
SDSSJ143759.21+382154.4&588017603622142042 &9.71&17.07 &100.0 &0.36&0.035& 0.16&	0.05		\\
SDSSJ144819.69+090702.1&588017702411763744& 9.54&22.04 &68.0& 3.13&0.029&0.19 &			0.03	\\
SDSSJ145146.60+523510.6&587733603184607338 &10.02&28.46 &263.0 &0.37& 0.065      & 0.54 &     0.48  \\
SDSSJ152819.60+530347.0& 588011102640930946& 10.16&27.77 &145.0& 0.30&0.053    &0.62 &0.60        \\
\hline
\end{tabular}
\caption{The main physical properties of the galaxy pairs
  sample. Stellar masses are calculated using the bulge+disk models
  from Mendel et al. (2014).  r$_{\mathrm{p}}$ and $\Delta$V represent
  the projected separation and the difference in velocity between the
  two members of the pair, respectively.  The mass ratio is calculated
  between the stellar mass of the galaxy which we observed with IRAM
  and that of its companion. Note that this ratio varies between
  $\sim$0.25 and 4 as we only selected potential major mergers. SFRs
  are from the MPA/JHU catalogue and estimated through the method
  presented in \citet{Brinchmann04}.  In the last column,
  aperture-corrected SFRs are reported, which represent the SFRs
  within the IRAM 30-m 22 arcsec beam and whose calculation is
  described in Section 4.2.}
\end{table*}

\section{Sample selection}
In order to investigate the effect of galaxy interactions on the molecular gas content of mergers
we selected a sample of galaxies with a close spectroscopic companion.
The parent sample is made up of more than 23000 galaxy pairs (Ellison et al. 2008, 2010, 2011; \citealt{Patton11}) from the Sloan Digital Sky Survey Data Release 7 (SDSS DR7, \citealt{Abazajian09}), to which we applied the following criteria.
Firstly, the galaxy must have a close companion at a projected separation r$_{\mathrm{p}}$ $\leq$30 kpc and the velocity separation between the two galaxies must be $\Delta$V $\leq$ 300 km s$^{-1}$; this latter condition maximises our chance of selecting true interacting systems rather than objects lying close in the sky as a result of projection effects. 
Based on these criteria, the selected galaxies are most likely caught either prior to the first encounter, or soon after the first pericentre passage.
Sources which reside at a more advanced phase in the merging event  (i.e. after the second pericentre passage) usually exhibit smaller separations and a more pronounced disturbed morphology  (e.g. \citealt{Renaud14}).  
Next, to strictly select objects undergoing a major merger, we also imposed the constraint that the companion's stellar mass must be within a factor of 4 of its own. Since our target galaxy is not necessarily the primary (most massive) of the pair, the mass ratios of our sample ranges between 0.25 and 4 (see Table 1).
To calculate stellar masses we used the bulge+disk models from Mendel et al. (2014).
Furthermore, the two galaxies which make up the pair must have a sufficiently large angular separation to avoid flux blending within the telescope beam.
The IRAM 30-m beam Full Width Half Maximum (FWHM) is $\simeq$11 arcsec at 2 mm, therefore we imposed the angular
separation of the pair to be at least $\geq$ 11 arcsec. 
In addition, a lower limit to the sSFR $\geq$ 3.9 Gyr$^{-1}$ was also imposed, so that the galaxies of the sample have relatively high SFRs for their mass, as expected due to the triggering in mergers.  In applying this cut we relied on SFR estimates reported in the MPA/JHU catalogue (http://wwwmpa.mpa-garching.mpg.de/SDSS/DR7), and calculated following \cite{Brinchmann04}.
Finally, in order to keep the exposure times to a reasonable value of $\leq$ 5 hrs per source, a mass cut of logM$_{*} \geq$ 9.5 M$_{\odot}$ was imposed. The final sample which satisfies all the above criteria is made up of a total of 12 sources.

In Figure 1, the SFR of our sample of pairs is compared with all star-forming galaxies, by computing an SFR offset ($\Delta$SFR).   
This comparison method, is analogous to our previous papers in this series, employed to determine differences in SFR, colour, metallicity,
HI content and AGN fraction in mergers (Ellison et al. 2010, 2011, 2013a, 2015, Patton et al. 2011, 2013, 
\citealt{Scudder12}, \citealt{Satyapal14}).
In brief, each galaxy in our pairs sample is matched in both redshift and stellar mass to a minimum of five control galaxies from the SDSS, with a nominal tolerance of 0.005 and 0.1 dex respectively. These tolerances are allowed to grow by 0.005 and 0.1 dex respectively until the minimum required number of control sources is reached. 
The `SFR offset' $\Delta$SFR, is defined as:
\begin{equation}
 \mathrm{\Delta SFR} = \mathrm{log(SFR^{tot},pair)} - \mathrm{log(SFR^{tot},control)}
\end{equation}

where $\mathrm{SFR^{tot},pair}$ and $\mathrm{SFR^{tot},control}$ are the
median total SFR of the galaxies in the pair and of the SDSS control sources, respectively.
The mean SFR offset of our galaxy pairs is 0.5 dex, as illustrated in the bottom panel of Figure 1.

\begin{table*}
\centering
\hspace{-0.2in}
\begin{tabular}{lccc} %K{5cm}K{2cm}K{2cm}}
%{l|r|r|l}
\hline
Source   & Tot. Int. Time & $\tau_{\mathrm{225 GHz}}$ & Number of scans  \cr 
&[mins] \cr  
\hline
SDSSJ014845.47+134300.2 & 163.8 & 0.084 & 17  \cr
SDSSJ080555.42+135959.0& 148.8 & 0.041 & 16\cr
SDSSJ111633.85+284606.4& 38.6 & 0.062 &4\cr
SDSSJ112036.59+361234.4& 123.2 & 0.081 & 13 \cr
SDSSJ123935.85+163516.1& 28.4 & 0.063& 3\cr
SDSSJ125053.09+352404.9&95.8 & 0.072& 10\cr
SDSSJ143154.09+215618.3& 18.8 & 0.201& 2\cr
SDSSJ143759.21+382154.4& 69.4 & 0.253& 9\cr
SDSSJ144819.69+090702.1& 58.6 & 0.092 &6\cr
SDSSJ145146.60+523510.6& 300.8 & 0.140 &32\cr
SDSSJ152819.60+530347.0& 125.6 & 0.044 &13\cr
\hline
\end{tabular}
\caption{Details of the IRAM 30-m observations of local galaxy pairs. Total integration time, average atmospheric opacity at 225 GHz ($\tau_{\mathrm{225 GHz}}$) and number of scans per source are reported. }
\end{table*}

\section{Observations and data reduction}	 

We observed 11/12 galaxies of our pairs sample with the IRAM 30-m Telescope at Pico Veleta (Spain), between the 15th and the 19th of December 2011,
under generally good weather conditions (0.04$\leq \tau_{225 GHz} \leq$0.25; where $\tau_{225 GHz}$ represents the optical thickness).
Our observing strategy aimed to achieve uniform $\geq$5$\sigma$ CO(1-0) and CO(2-1) line-peak sensitivity across the whole sample, thus the integration time spent on each source varied between $\sim$18 and $\sim$300 minutes (see Table 2).
The Eight Mixer Receiver (EMIR; \citealt{Carter12}) was used, which is characterized by two side bands of 8 GHz width each and two polarizations.
Dual band observations with the combination E0(3 mm)--E2(2 mm) were performed in order to observe the CO(1--0) and CO(2--1) lines simultaneously. Our galaxy mergers span the range 0.023$\leq$ z $\leq$0.065 with CO(1--0) and CO(2--1) redshifted between 108.210--112.675 GHz and 216.417--225.345 GHz respectively.
We therefore set up the E2 receiver to cover the CO(2--1) line with three different tunings: 217.775, 221.667 and 224.283 GHz. The E0 receiver was correspondingly tuned to 108.889 GHz for the first E2 setup and 111.503 for the other two.
The Wideband Line Multiple Autocorrelator (WILMA) was used as the back-end: it covers 4 GHz in each linear polarization for each band and gives a resolution of 2 MHz. As backup, the data were also recorded by the Fast Fourier Transform Spectrometers (FTS).
Due to poor weather conditions one source could not be observed, therefore our study is based on a sample of 11 objects.
In the left panels of Fig. 2 we show the SDSS cutouts of the 11 interacting systems, with the galaxies we observed encircled in white with the beam size of the IRAM 30-m telescope at 3 mm.

The data reduction was carried out with the  Continuum and Line Analysis Single-dish Software (\textit{CLASS}; http://www.iram.fr/IRAMFR/GILDAS); hereafter we describe the standard reduction procedure for both CO (1--0) and CO(2--1) spectra, which is, for consistency, the same one adopted by \cite{Saintonge11a} for the COLDGASS spectra.  
All the scans were visually examined, and those with severe baseline issues were rejected.  
The baseline of each scan was then fitted with a first-order polynomial and subtracted. All the scans belonging to the same observed galaxy are then combined together to generate an average spectrum which is later smoothed to a resolution of 20 km s$^{-1}$, yielding a 1$\sigma$ channel rms 
of $\sim$1.5 mK.
The total emission line flux I$_{\mathrm{CO}}$ is obtained by integrating the signal 
within a manually defined spectral window spanning the FWHM of the line. \\

The final spectra are shown in Figures 2 and 3, and the results of our reduction are presented in Table 3.
The formal errors of the integrated line fluxes I$_{CO}$ were calculated as:

\begin{equation}
 \sigma_{I}= \frac{\sigma_{rms} W50_{CO}} {({W50_{CO} \Delta w^{-1}})^{0.5}} 
\end{equation}

Where $\sigma_{\mathrm{rms}}$ is the rms noise per spectral channel of width $\Delta w_{\mathrm{channel}}$=21.57 km s$^{-1}$ and W50$_{\mathrm{CO}}$ is the line width calculated as in \cite{Saintonge11a}.

\begin{table*}
\centering
\hspace{-0.2in}
\begin{tabular}{ccccccc} %K{5cm}K{2cm}K{2cm}K{2cm}K{2cm}K{2cm}K{2cm}}
\hline
Source & I$_{\mathrm{CO(1-0)}}$ & W50$_{\mathrm{CO(1-0}})$ & L$^{\prime}_{\mathrm{CO(1-0}})$ & I$_{\mathrm{CO(2-1}})$ &  W50$_{\mathrm{CO(2-1}})$ & L$^{\prime}_{\mathrm{CO(2-1}})$  \\
 &[Jy km s$^{-1}$] & [km s$^{-1}$] & [10$^{8}$K km s$^{-1}$ pc$^{2}$] & [Jy km s$^{-1}$] & [km s$^{-1}$] & [10$^{8}$K km s$^{-1}$ pc$^{2}$] \\
\hline
SDSSJ014845.47+134300.2 & 1.89$\pm$ 0.05 & 164 & 1.79 $\pm$ 0.18 & 4.53$\pm$ 0.05 & 153 & 1.05 $\pm$ 0.11\\
SDSSJ080555.42+135959.0 & 5.24$\pm$ 0.06  & 259 &3.49$\pm$ 0.35  &10.35 $\pm$ 0.07 & 239 & 1.76$\pm$ 0.18 \\
SDSSJ111633.85+284606.4 & 4.20$\pm$ 0.04 & 29 &1.06 $\pm$ 0.11 & 5.99 $\pm$ 0.05 & 33 &0.38$\pm$  0.09\\
SDSSJ112036.59+361234.4 & 3.25$\pm$ 0.07 &270&4.07$\pm$ 0.42&6.52$\pm$  0.11 &209 &2.06$\pm$ 0.21\\
SDSSJ123935.85+163516.1 & 4.79$\pm$ 0.13 & 154&1.48$\pm$ 0.15  &7.58 $\pm$ 0.13 & 165 & 0.59$\pm$ 0.06 \\
SDSSJ125053.09+352404.9 & 2.14$\pm$ 0.05 &166&1.06$\pm$ 0.11 &4.73$\pm$  0.11 &162&0.59$\pm$ 0.06\\
SDSSJ143154.09+215618.3 & 27.21$\pm$ 0.21 &200 &24.89$\pm$  2.50 &32.29 $\pm$ 0.26 &190 &7.34$\pm$ 0.74\\
SDSSJ143759.21+382154.4  & 3.81$\pm$ 0.10 &208 & 2.20$\pm$ 0.23 & 7.02 $\pm$ 0.12 & 199 &1.01$\pm$ 0.10\\
SDSSJ144819.69+090702.1 & 1.81$\pm$ 0.08 &115&0.72$\pm$ 0.08 &2.28$\pm$ 0.124 &18 &0.23$\pm$ 0.03\\
SDSSJ145146.60+523510.6 & 1.93$\pm$ 0.03 &  178 &3.85 $\pm$ 0.39 &4.12 $\pm$ 0.06 &167 &2.06$\pm$  0.21\\
SDSSJ152819.60+530347.0 & 6.67$\pm$  0.05 & 183& 8.81$\pm$ 0.88&11.96$\pm$ 0.07 &180 &3.96 $\pm$ 0.40\\
\hline
\end{tabular}
\caption{Results from our CO (1--0) and (2--1) observations of our sample of 11 local mergers. CO emission line intensity (I$_{\mathrm{CO}}$), width (W50$_{\mathrm{CO}}$) and luminosity (L$^{\prime}_{\mathrm{CO}}$) are reported for each transition. The method to calculate these quantities is described in Section 3 and the corresponding spectra are shown in Figure 2 and 3.}
\end{table*}

\begin{table*}
\centering
\hspace{-0.2in}
\begin{tabular}{cccccc} %K{2cm}K{5cm}K{2cm}K{2cm}K{2cm}K{2cm}}
\hline
Source& f$_{SB}$&$\alpha_{\mathrm{CO}}$  & $\log M_{\mathrm{H_2}}$  & f$_{\mathrm{gas}}$ & t$_{\mathrm{dep}}$  \\
& & [M$_{\odot}$ (K km s$^{-1}$ pc$^{2}$)$^{-1}$]   &[M$_{\odot}$]  & & [Gyr] S \\
\hline
SDSSJ014845.47+134300.2& 0.25&4.25  & 8.88&0.14& 0.42		\\
SDSSJ080555.42+135959.0&0.19 &5.03 &  9.24  &0.28 &	0.71\\
SDSSJ111633.85+284606.4&0.05  &7.60&  8.91 &0.27 &	1.04\\
SDSSJ112036.59+361234.4&0.11 &4.43 & 9.26 &0.18 & 0.63\\
SDSSJ123935.85+163516.1& 0.08 &3.65 & 8.73 & 0.10 &	0.44\\
SDSSJ125053.09+352404.9& 0.03 &4.20 &  8.65&0.11 &	0.42\\
SDSSJ143154.09+215618.3& 1.00 &0.97 & 9.38 & 0.06 & 0.05		\\
SDSSJ143759.21+382154.4&0.05  &4.49 &8.99 &0.19 &	0.88	\\
SDSSJ144819.69+090702.1& 0.14 &4.99 & 8.55&0.10 &0.34	\\
SDSSJ145146.60+523510.6&0.06  &4.51 &9.24 &0.16 & 0.58		\\
SDSSJ152819.60+530347.0&0.11 &3.54 & 9.49&0.22 &	0.79\\
\hline
\end{tabular}
\caption{Derived physical quantities of our sample of 11 local
  galaxies in pairs.  The value $f_{\mathrm{SB}}$ is the probability
  of a pair galaxy to be in a starburst phase given its position in
  the SFR--M$_{*}$ plane (see section 4.1).  The conversion factor
  $\alpha_{\mathrm{CO}}$ has been calculated following
  \citet{Sargent14} and described in section 4.1; H$_{2}$ masses are
  derived from the CO(1--0) transitions. Molecular gas fractions are
  calculated as f$_{gas}=$M$_{\mathrm{H_{2}}}$/M$_{*}$ and H$_{2}$
  depletion times are
  t$_{\mathrm{dep}}$=M$_{\mathrm{H_{2}}}$/SFR$^{\mathrm{aperture}}$. The
  average errors on gas masses, gas fractions and depletion times are
  10$\%$, 42$\%$ and 14$\%$, respectively (see the text for a
  description of their calculation). }
\end{table*}
\section{Analysis and Results}

\subsection{CO luminosities and molecular gas masses}
To compute the molecular gas content of our targets we
calculate the CO line luminosities using the following equation (\citealt{Solomon97}):

\begin{equation}
L_{\mathrm{CO}}^{\prime}=3.25 \times 10^{7} I_{\mathrm{CO}} \nu_{\mathrm{obs}}^{-2} D_{\mathrm{L}^{2}} (1+z)^{-3}
\end{equation}

where L$^{\prime}_{\mathrm{CO}}$ is the CO luminosity in K km s$^{-1}$ pc$^{2}$,  I$_{\mathrm{CO}}$ represents the line flux in units of Jy km s$^{-1}$,  $\nu_{\mathrm{obs}}$ is the observed frequency of the line units of GHz and D$_{L}$ is the luminosity distance expressed in Mpc (see Table 3).
In the following analysis we only utilize the luminosity from the CO(1--0) transition, as this is the best tracer of the total molecular gas reservoir.
The CO(2--1) transition, in fact, traces the gas which is in a slightly denser phase (e.g. \citealt{Solomon05}), and therefore provides a less accurate estimate of the total molecular gas reservoir. In addition, the IRAM beam size at 2 mm is characterized by a FWHM of 11 arcsec, and is therefore sensitive to the gas which only resides in the innermost part of the galaxy ($\leq$8 kpc).
Excitation in the central region of the galaxies as traced by the I$_{\mathrm{CO(2-1)}}$/I$_{\mathrm{CO(1-0)}}$ ratio will be the subject of future work. 
The molecular hydrogen (H$_{\mathrm{2}}$) masses within the IRAM 22 arcsec beam can be computed as M$_{\mathrm{H_{2}}}^{\mathrm{ap}}$=L$^{\prime}_{\mathrm{CO}}$ $\times$ $\alpha_{\mathrm{CO}}$ (in the rest of the paper we refer to M$\mathrm{^{ap}_{H_{2}}}$ as simply M$_{\mathrm{H_{2}}}$).
We compute CO-to-gas conversion factors $\alpha_{\mathrm{CO}}$ on a per-galaxy basis following the "2-Star Formation Mode (SFM)" framework of Sargent et al. (2014). Specifically, $\alpha_{\mathrm{CO}}$ values are calculated as:
\begin{equation}
 \alpha_{\mathrm{CO}} = (1-f_{\mathrm{SB}})\times \alpha_{\mathrm{CO,MS}} + f_{\mathrm{SB}}\times \alpha_{\mathrm{CO,SB}} 
\end{equation}

where  $f_{\mathrm{SB}}$ is the probability of a galaxy being in a starburst phase given its
offset from the mean locus of the star-forming main sequence in the SFR--M$_{*}$ plane (\citealt{Sargent12}), while $\alpha_{\mathrm{CO,MS}}$ and $\alpha_{\mathrm{CO,SB}}$ are
the CO-to-H$_{2}$ conversion factors expected in the 2-SFM formalism for a galaxy with the SFR and M$_{*}$ values determined for a given galaxy in a pair. 
The main-sequence value $\alpha_{\mathrm{CO,MS}}$ varies with the galaxy metallicity following the \cite{Wolfire10} prescription. $\alpha_{\mathrm{CO,SB}}$
deviates from the MS-$\alpha_{\mathrm{CO}}$ by an amount which depends on the intensity of the starburst (i.e. on the sSFR offset from the main sequence, see Sargent et al. (2014) for a full description 
of the underlying calculations). 
For this sample, metallicities are taken from \cite{Tremonti04}.

In Table 4 we report the $f_{\mathrm{SB}}$  and $\alpha_{\mathrm{CO}}$ values estimated for each of our sources together with the H$_\mathrm{{2}}$ masses.
The conversion factor $\alpha_{\mathrm{CO}}$ varies between 0.97 and 7.60, with a median value of 2.29 M$_{\odot}$ (K km s$^{-1}$ pc$^{2}$)$^{-1}$, i.e. about a factor of $\sim$2 lower than
 the canonical MW-conversion factor and reflecting the fact that galaxies in our sample are offset to high sSFRs. 

Molecular gas masses span the range 8.5$\leq$ log(M$_{*}$/M$_{\odot}$) $\leq$9.5
with a mean value of log(M$_{*}$/M$_{\odot}$)=9.12.
Two factors contribute to the uncertainties on the H$_{\mathrm{2}}$ masses:
 the error on the integrated line intensity $\sigma_{\mathrm{I}}$ ($\leq$2$\%$) and a flux calibration error which is $\sim$10$\%$ for 3 mm observations (\Citealt{Saintonge11a}). The total average error is consequently $\sim$10$\%$.
We do not include redshift uncertainties (which are negligible compared to the flux uncertainties), nor the systematic uncertainties involved in the calculation of  $\alpha_{\mathrm{CO}}$ values (as these will affect conversion factors estimated for our control sample in exactly the same way).

\begin{figure}
\centering
\includegraphics[width=90mm,angle=0]{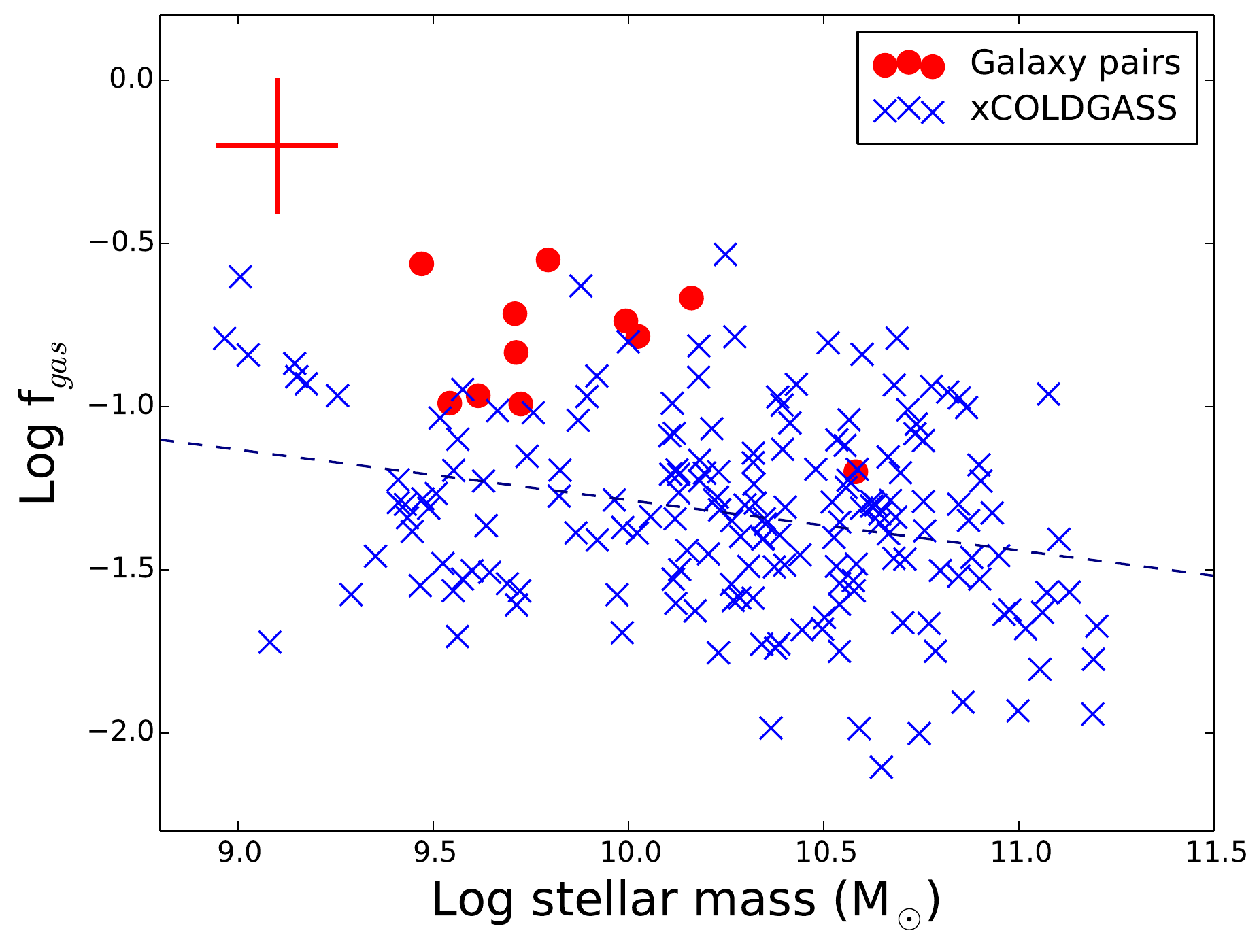}
\caption{Molecular gas fraction f$_{gas}$=M$_{H_{2}}$/M$_{*}$ plotted as a function of the stellar mass M$_{*}$ for our 11 galaxies in pairs (filled red circle) and xCOLDGASS sources (blue crosses). The red cross
represents the average errors on gas fraction and stellar mass of our sample. Galaxy pairs displays enhanced H$_{2}$ content, and lie on average $\sim$0.44 dex ($\sim$1.3$\sigma$) above  the dashed blue line which represents the best linear fit to xCOLDGASS galaxies.}
\end{figure}

\subsection{Aperture SFRs}

The FWHM of the IRAM 30-m telescope is 11 and 22 arcsec at 2 and 3 mm, respectively, corresponding to $\sim$8 and 16 kpc at the median redshift of our targets. For galaxies with extended gas distributions, CO flux from the outer part of the galaxy could thus potentially be missed by single-pointing observations (see Fig. 2).
The most common approach to deal with this issue is to apply an aperture correction to the CO flux measurements. This is frequently done by utilising resolved CO maps of similar sources (\citealt{Regan01}; \citealt{Kuno07}; \citealt{Leroy09}) to estimate the amount of missed flux (\citealt{Bothwell13}; \citealt{Saintonge11a}), assuming a smooth gas profile
which follows the distribution of stellar light.
However, studies based on both isothermal simulations and spatially resolved maps show that during a galaxy encounter the distribution of molecular gas is less uniform, with CO emission located along tidal features, dust lanes or in extended disks (\citealt{Shlosman89}; \citealt{Konig14}; \citealt{Ueda14}). 
For this reason, we adopt a slightly different approach to derive aperture corrections.
Instead of applying a correction to the CO flux, to compute depletion times we have adjusted the SFRs in order to estimate
the value of this quantity only within a 22 arcsec beam.
%Initial stellar masses are derived from r-band photometry s described in \cite{Mendel14}.
For our sample, SDSS total SFRs are available from the MPA/JHU catalogue.
These have been calculated either by modelling the emission lines with the \cite{Charlot01} models, or, if the galaxy hosts an AGN according to the \cite{Kauffmann03} classification, by using the SFR--D4000 relation (\citealt{Brinchmann04}). We thus calculate the fraction of the total r-band flux emitted within the 22 arcsec IRAM beam size and we multiply it by the total SFR from the MPA/JHU catalogue to obtain an aperture-converted SFR.
%For consistency, the same method has been adopted in deriving aperture-corrected SFRs of our comparison sample, i.e. xCOLDGASS, which is described in Section 4.2.

\subsection{Comparison sample: xCOLDGASS }

In order to robustly compare the molecular gas and the star-forming properties of our galaxy pairs with non-interacting galaxies, we require
a carefully constructed comparison sample of `normal' star-forming galaxies.
The extended  CO Legacy Data base for the GALEX Arecibo SDSS Survey
(xCOLDGASS, \cite{Saintonge17}) represents the ideal sample for this purpose.
COLDGASS (Saintonge et al. 2001a, 2001b, 2012)
 is a legacy survey  which studies the molecular gas of nearby late-type galaxies through IRAM 30-m CO(1--0) and CO(2--1) 
observations.
It is comprised of 365 SDSS sources
in the redshift range 0.0025 $\leq$z $\leq$0.05  with stellar masses  10$\leq$ log(M$_{*}$/M$_{\odot}$) $\leq$11.5.
xCOLDGASS is an extension which also includes sources with  masses down to log(M$_{*}$/M$_{\odot}$)=9, bringing the total to 500 sources.
Out of this extended sample, we only use galaxies with CO(1--0) detections and we additionally required a stellar mass measured by Mendel et al. (2014), which leaves 270 sources. 
Finally, in order to ensure that our control sample is made up of only `isolated' star-forming galaxies, we excluded 
all those galaxies which have either a spectroscopic companion within 80 kpc and $\Delta$V$\leq$300 km s$^{-1}$, or have 
a Galaxy Zoo merger vote fraction $\geq$0 (See Darg et al., 2010 for further details on this last criterion).  
The final xCOLDGASS sub-sample we take into consideration is thus formed of 186 galaxies. 
SDSS SFRs of xCOLDGASS galaxies have been aperture-corrected following the method described in Section 4.2.
For consistency with our own measurements, H$_{2}$ masses are calculated by considering only the CO emission within the 22 arcsec
IRAM beam, and by multiplying L$^{\prime}_{CO}$ by an $\alpha_{\mathrm{CO}}$ conversion factor which varies for each source estimated using the method presented in Section 4.1.
To summarize, the xCOLDGASS sub-sample we are employing in our analysis has stellar masses, SFRs and molecular gas masses
 calculated  with exactly the same techniques as our 11 galaxy pairs, permitting a robust and like-for-like comparison.

\begin{figure}
\centering
\includegraphics[width=90mm,angle=0]{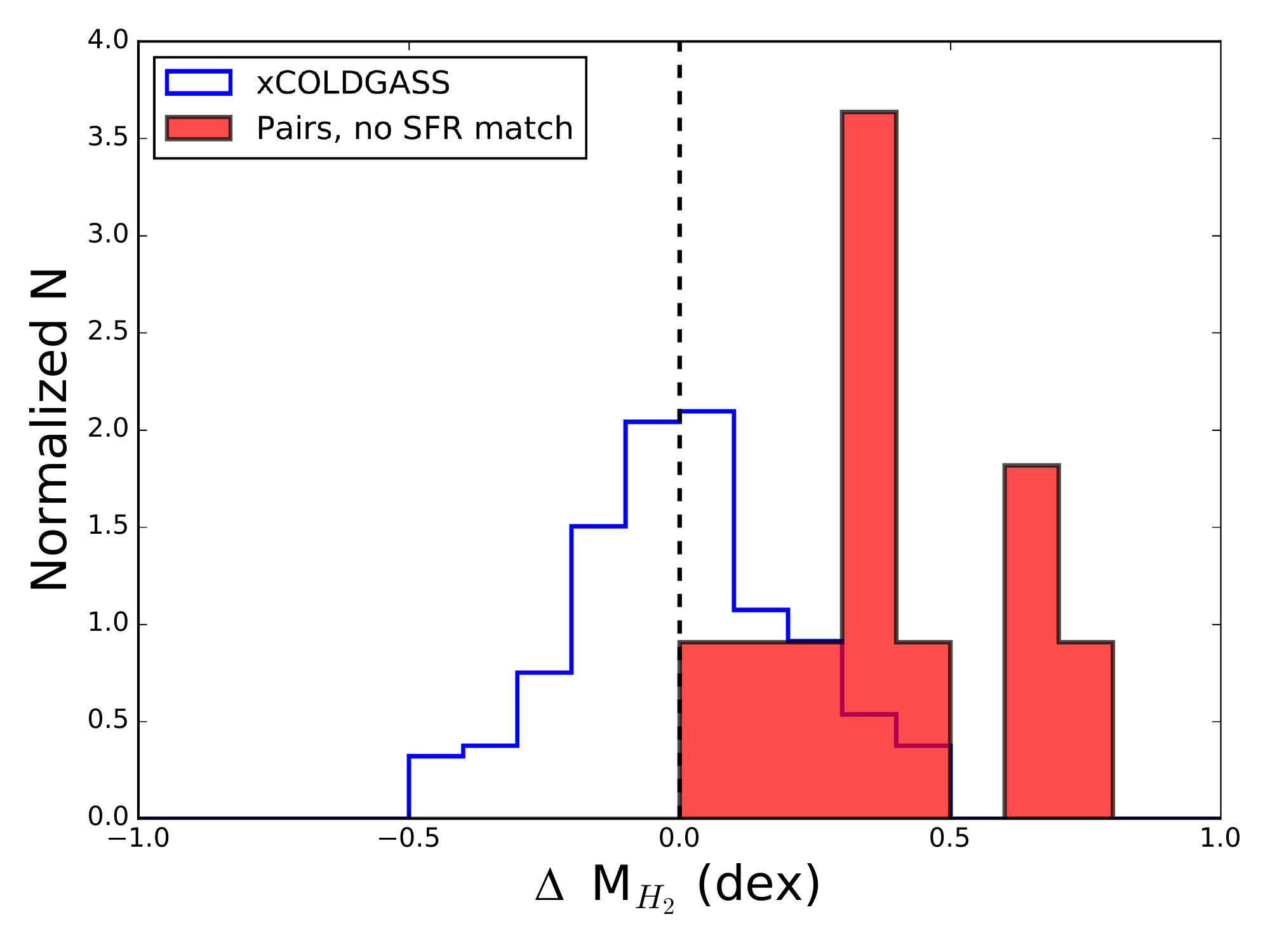}
\includegraphics[width=90mm,angle=0]{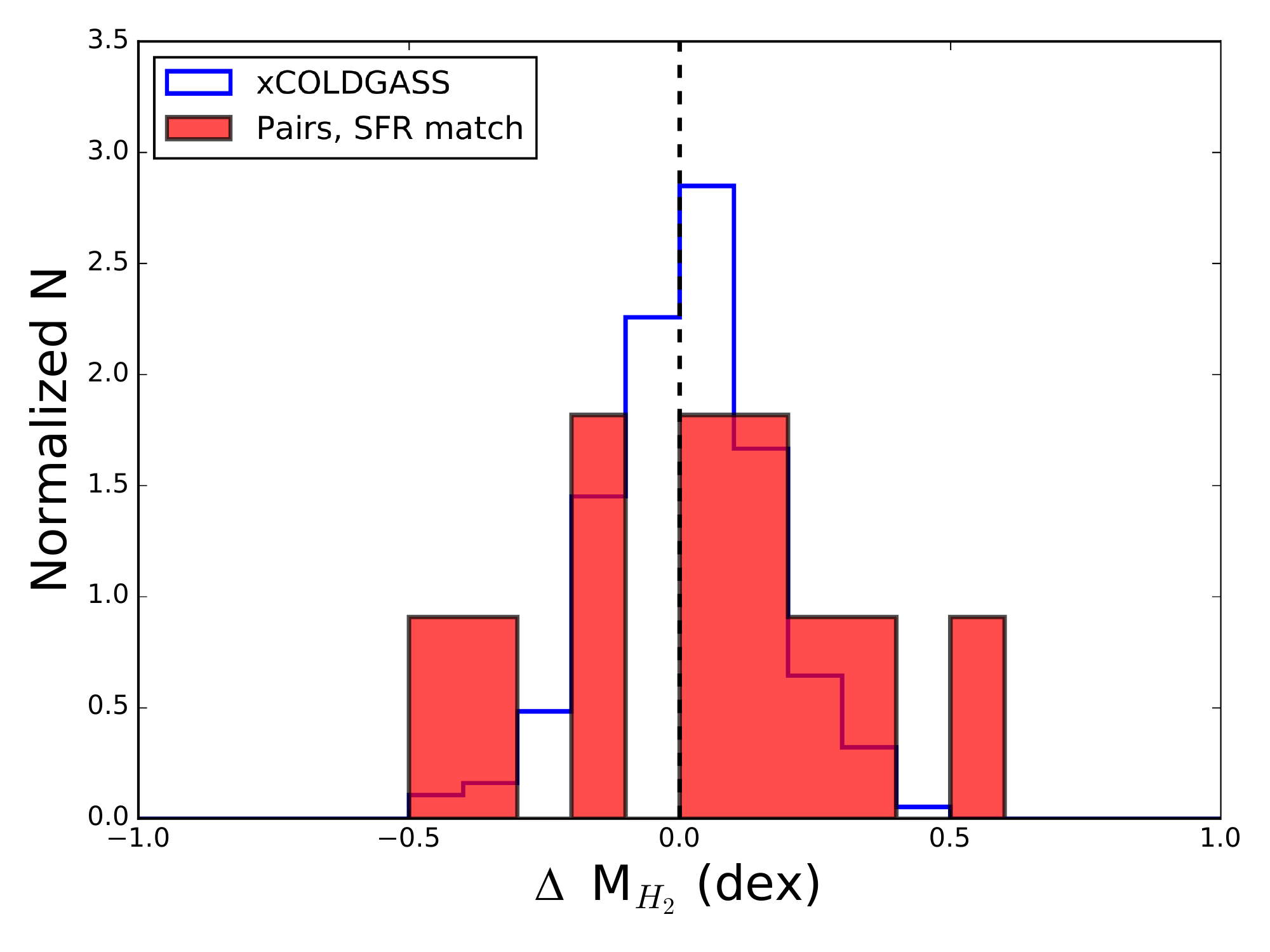}
\caption{Distribution of molecular gas mass offset. Our sample of galaxy pairs is represented by the filled red histogram, whereas the xCOLDGASS sample distribution is plotted as a blue histogram. \\Top panel: $\Delta$M$_{H_{2}}$ calculated employing as matching parameters stellar mass, redshift and local density. Our pairs sample has a median M$_{H_{2}}$ offset of 0.34 dex with respect to the control sample.\\
Bottom panel: Same as above, with the addition of the SFR as matching parameters. The median offset $\Delta$M$_{H_{2}}$ of the galaxy pairs is 0.07.}
\end{figure}

\begin{figure}
\centering
\includegraphics[width=90mm,angle=0]{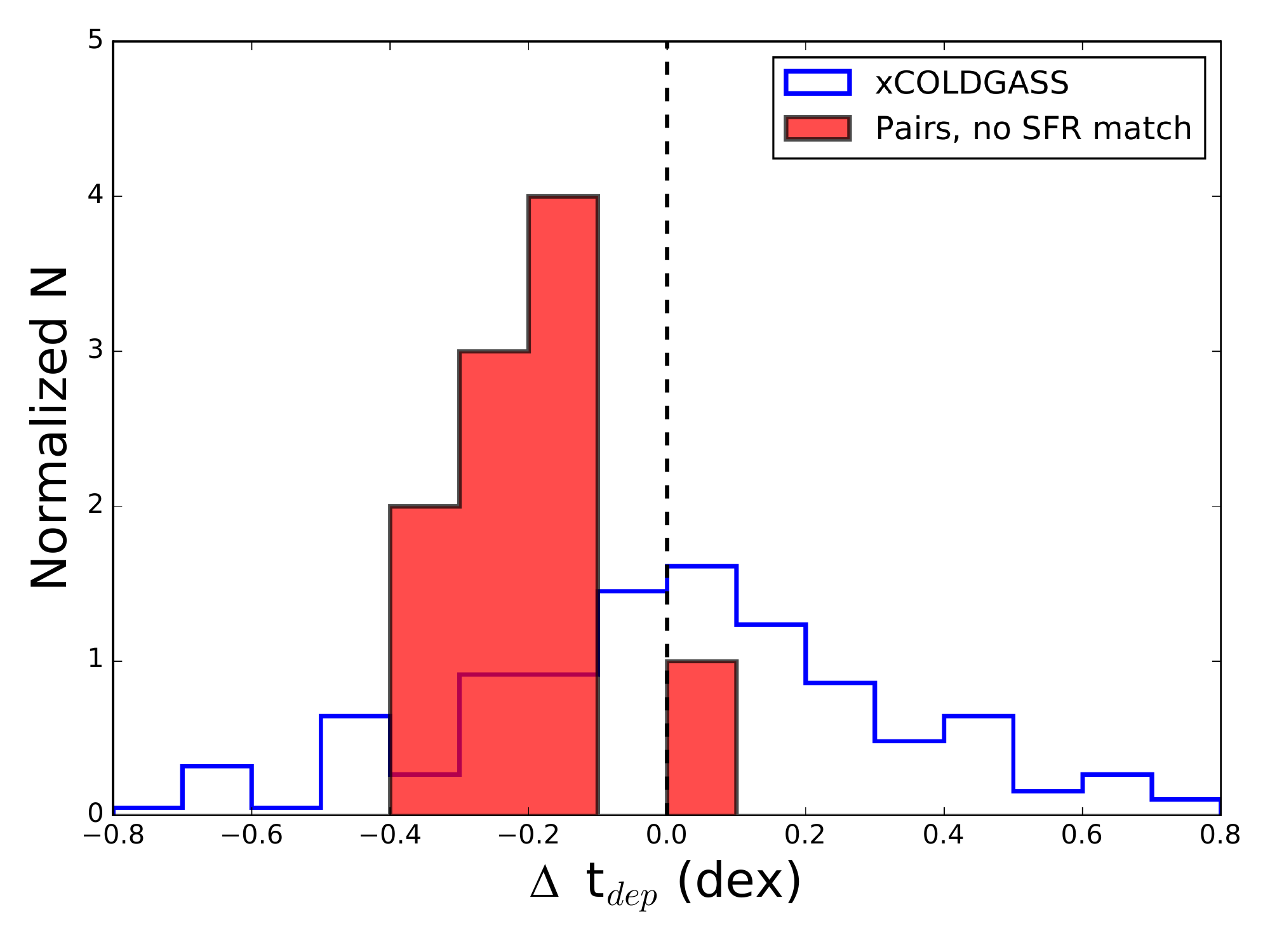}
\includegraphics[width=90mm,angle=0]{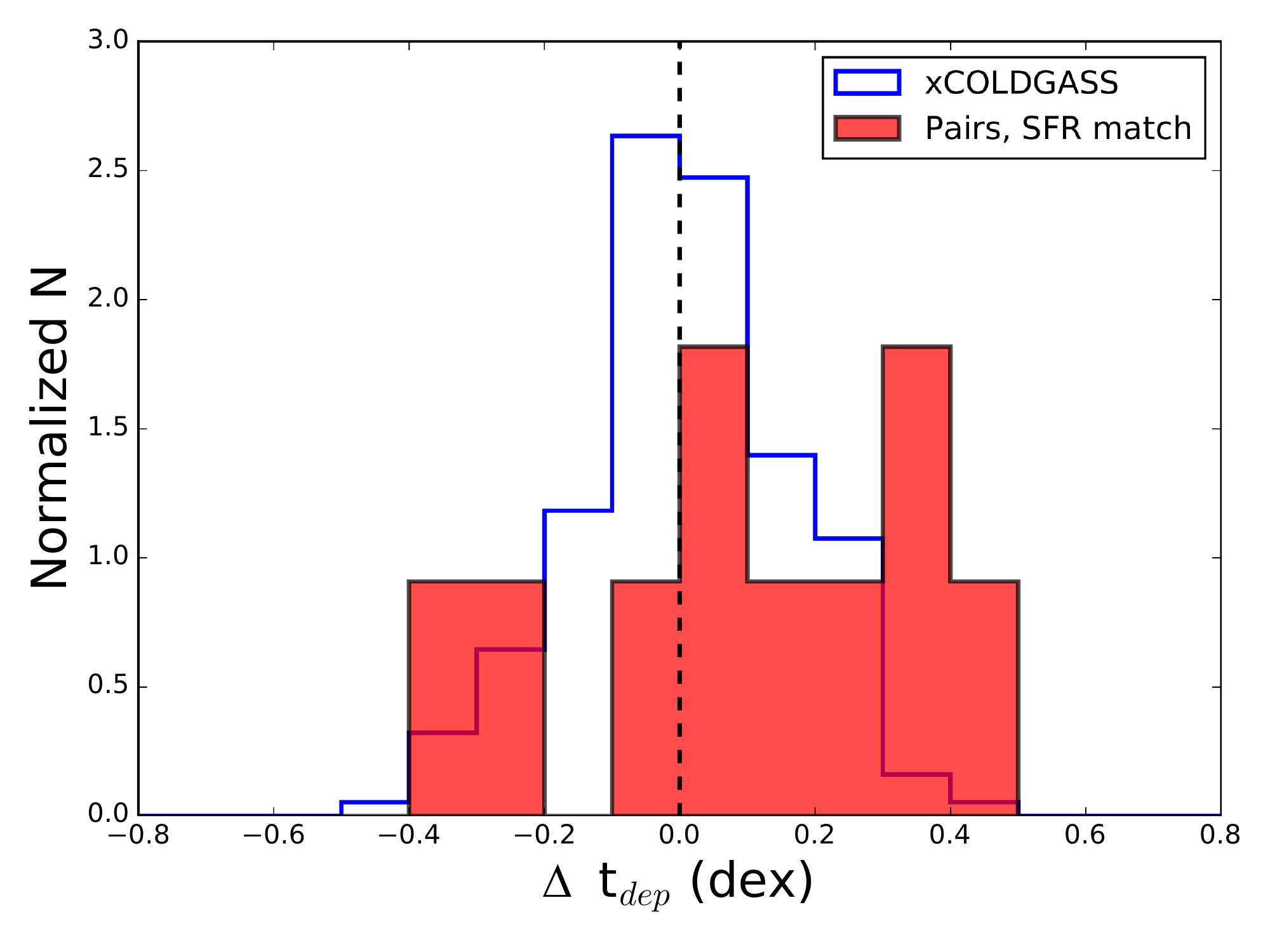}
\caption{Molecular gas depletion time offset distribution of galaxy pairs (red filled histogram) and xCOLDGASS sources (blue empty histogram). \\Top panel: Comparison carried out using the following matching parameters: M$_{*}$, z and $\delta_{5}$. Our sample has a median offset from the matched control sample of -0.21 dex. \\
Bottom Panel: Same as above, with the inclusion of the SFR in the matching procedure; the median $\Delta$t$_{\mathrm{dep}}$ is 0.09. }
\end{figure}

\subsection{Molecular gas masses}

As a first step in our analysis, we compare the H$_{\mathrm{2}}$ content in the 11 galaxy pairs with those of normal star-forming galaxies.
Molecular gas fractions (f$_{\mathrm{gas}}=$M$_{\mathrm{H_{2}}}$/M$_{*}$) for our galaxy pairs are presented in Table 3. They vary from 0.06 $\leq$ f$_{gas} \leq$ 0.26 with a median value of 0.16$\pm$0.03. The average errors on the gas fractions are $\sim$42$\%$, taking into account both the errors on the H$_{2}$ masses and those on the stellar masses, which are on average $\sim$40$\%$ (\citealt{Mendel14}). 
 
It is worth noting that the gas fractions obtained likely represent lower limits, since we are not including CO emission outside the IRAM 22 arcsec beam (which corresponds to a physical extent of $\sim$16 kpc). However, we also use this approach for the control sample, to ensure that 
the comparison analysis remains consistent. 
In Figure 4 we plot the molecular gas fraction as a function of the stellar mass for both the 11 galaxies in pairs and the 186 sources from xCOLDGASS for comparison. The blue dashed line is the best linear fit to the xCOLDGASS galaxies: as expected, the gas fraction decreases with increasing stellar mass, as found in numerous previous studies (e.g. \citealt{Bothwell14}; \citealt{Saintonge11a}).
All the galaxies in pairs have higher gas content compared to normal galaxies, lying  $\sim$0.4 dex above the general trend defined by the entire xCOLDGASS sample. 
To assess in a more statistical way the difference between the relative H$_{2}$ content of galaxy pairs with respect to the xCOLDGASS sample we perform a Kolmogorov-Smirnov (KS)
test, which can indicate whether or not two populations are drawn from the same underlying distribution of the gas fraction. The test returns a statistic of D=0.82 with a p-value=4.125$\times$10$^{-7}$, which corresponds to  $\geq$99.99$\%$ probability that the molecular gas fractions of our 11 interacting galaxies and those of normal galaxies belong to two intrinsically distinct distributions. 
 
We now seek to compare each of our merging galaxies only with those sources from xCOLDGASS which exhibit similar underlying physical properties.
We compute $\Delta$M$_\mathrm{{H_{2}}}$, which
quantifies the discrepancy in the gas mass between interacting and isolated star-forming galaxies which lie in the same ranges of stellar mass, redshift and local environment, quantified by the parameter $\delta_{5}$. This quantity represents the local density, and is related to the distance of the galaxy neighbours
(see \citealt{Ellison11} for a full description).
The quantity $\Delta$M$_{\mathrm{H_{2}}}$ is analogous to $\Delta$SFR presented in Section 2, and is defined as:

\begin{equation}
 \Delta M_{\mathrm{H_{2}}} = log(M\mathrm{_{H_{2}}},pair) - log(M\mathrm{_{H_{2}}},control)
\end{equation}

where log(M$\mathrm{_{H_{2}}},pair$)  and log(M$\mathrm{_{H_{2}}},control$) are the
molecular gas masses of the galaxies in the pair and the mean gas fraction of the control sources, respectively.
In carrying out this test we adopted matching tolerances in all parameters of 0.005 dex and allowed 2 control sources per paired galaxy.

The top panel in Figure 5 shows the distribution of $\Delta$M$\mathrm{_{H_{2}}}$ of the 11 galaxies in pairs compared to that of galaxies from
xCOLDGASS. The median $\Delta$M$\mathrm{_{H_{2}}}$ of our sample is 0.34 dex, confirming the results from Fig. 4, and a KS test indicates that
the $\Delta$M$\mathrm{_{H_{2}}}$ of our sample of galaxy pairs and that of the xCOLDGASS sample are drawn from two different distributions with a probability $\geq$99$\%$ (p-value=3.33$\times$10$^{-5}$).
In order to test the effect of measurement errors in M$\mathrm{_{H_{2}}}$ on our calculation,
for each galaxy of the pairs and xCOLDGASS samples we deviate the value of M$_{\mathrm{H_{2}}}$ by an amount sampled from
within its gaussian uncertainty, thus generating two artificial samples of galaxy pairs and isolated galaxies.  
We repeat this procedure 10000 times, and for each iteration we perform a KS test between the $\Delta$M$_{H_{2}}$ of the two artificial samples,
and we register the number of times in which the two samples are statistically different at $\geq$3$\sigma$ level (i.e. p-value $\leq$0.003). We find that the pairs and the xCOLDGASS sample belong to two different underlying distributions for all the 10000 iterations. 

Although our results indicate that galaxies in pairs have a higher
molecular gas mass than non-interacting galaxies by about a factor
of two, we know that their SFRs are also enhanced (Fig. 1).
In order to take into account the enhanced SFR of our paired galaxies, we repeat the analysis described above by adding an extra matching parameter, i.e. SFR, for which we apply the same tolerance of the other parameters (0.005 dex). 
In the bottom panel in Figure 5 we show the new distribution of $\Delta$M$\mathrm{_{H_{2}}}$ of the 11 pairs sample and xCOLDGASS. The median $\Delta$M$\mathrm{_{H_{2}}}$ 
 of the sample of galaxy pairs drops to 0.07 and a KS test between $\Delta$M$\mathrm{_{H_{2}}}$ of the pairs and `normal' galaxies returns a p-value=0.36, indicating that the two sample are statistically indistinguishable. 
A bootstrap test shows that this result holds for the whole 10000 iterations in which the M$_{\mathrm{_{H_{2}}}}$ of galaxy pairs and xCOLDGASS galaxies are randomly resampled.

%This result is coherent with what is found by Sargent et al. (in prep.),
%who compared the molecular gas content of a sample of 
%local post-merger galaxies with that of a sample of normal isolated galaxies. Their results indicate that post-mergers have increased H$_{2}$ fraction by 0.74$\pm$0.19 dex wi%th respect to a control sample matched in stellar mass and redshift. However, differently from what is observed for our galaxy pairs, this enhancement is still statistically %significant (0.43$\pm$0.03 dex) once SFRs is also included as matching criteria, suggesting that
%post-merger galaxies exhibit intrisically higher molecular gas content.  

To summarize, although galaxy pairs show molecular gas mass enhancement 
relative to a stellar mass matched control,
once we account for the relatively high sSFR of our sample, the difference in molecular gas content between galaxy pairs and xCOLDGASS sources seems to disappear. However, a caveat to this analysis is represented by the small number of xCOLDGASS sources capable of matching the pairs in all the parameters considered, particularly in SFR.
In the future, a larger sample of control galaxies will be helpful in confirming this result.

\subsection{Molecular gas depletion times}

As demonstrated in Figure 1, our galaxy pair sample is selected to exhibit enhanced (total) SFRs with respect to main sequence galaxies.
To investigate whether or not this enhanced SFR is entirely a result of the larger H$_{2}$ mass providing additional fuel for star formation
we now directly study the molecular gas depletion time, defined as t$_{\mathrm{dep}}$=M$_{\mathrm{H_{2}}}$/SFR$^{\mathrm{aperture}}$, which is (by definition) the inverse of the star formation efficiency (SFE). The depletion time represents the length of time which is necessary to consume the whole gas reservoir of a galaxy by converting it into new stars, assuming a constant rate of star formation and
no replenishment of the gas reservoir nor any gas outflows.

\begin{figure}
\centering
\includegraphics[width=90mm,angle=0]{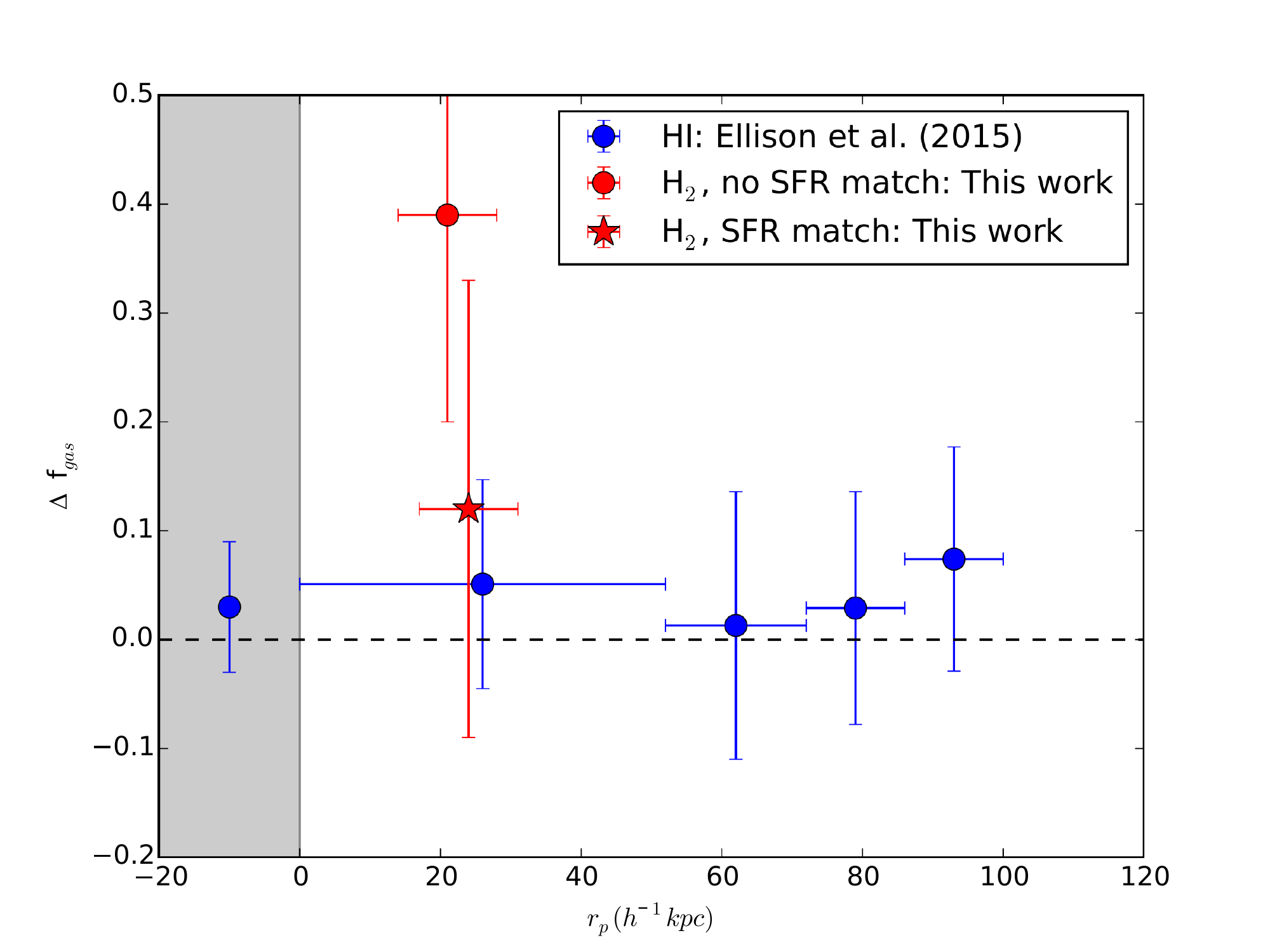}
\caption{Distribution of the mean neutral gas fraction offset
  ($\Delta$f$_{gas}$) as a function of projected separation. The blue
  circles are the pairs and post-merger galaxies from
  \citep{Ellison15}, whereas the red symbols represent the mean H$_{2}$
  fraction offsets of the 11 galaxy pairs of this paper (red points
  are slightly offset on the x-axis for clarity). The method employed
  to calculated the H$_{2}$ fraction offset is identical to that
  described in section 4.4.  }
\end{figure}

For the pairs sample we found the depletion time spans the range 0.05--1.04 Gyr with a median value of 0.57$\pm$0.1 Gyr. 
The depletion time of the full xCOLDGASS galaxies varies between $\sim$0.14 and 110 Gyr with a median value of 1.7$\pm$0.93 Gyr, approximately a factor of 2--3 longer than in the galaxy
pairs. A KS test reveals D=0.62 (p-value=2.2$\times$10$^{-4}$), which suggests that the depletion times of the 11 pairs and those of normal galaxies belong to two different distributions at a confidence level of $\gg$99.99$\%$ .
For a more meaningful comparison of the gas consumption timescales of interacting and normal galaxies, we again adopt the same technique used for the SFR and molecular gas masses comparison (i.e. we calculate a `depletion time offset'). 
The procedure adopted here is identical to that described in Section 4.4, meaning that our 11 galaxies in pairs are first compared to two control sources from xCOLDGASS with matched redshift, stellar mass and local density, and consequently the SFR is introduced as extra matching parameter (we keep the same tolerance adopted before, i.e. 0.005). 
The quantity we consider here is the 'depletion time offset', $\Delta$t$_{\mathrm{dep}}$, which is therefore:
\begin{equation}
 \Delta t_{\mathrm{dep}} = log(t_\mathrm{dep},pair) - log(t_\mathrm{{dep}},control)
\end{equation}
The histograms in Figure 6 show the $\Delta$ t$_\mathrm{{dep}}$ distributions of our sample and that of xCOLDGASS.
When the SFR is not included in the matching parameters, the median `depletion time offset' of galaxies in pairs is $-$0.21 dex (top panel),
indicating a depletion in the pairs that is 60 per cent shorter than in the control.
A KS test produces a p-value=5.8$\times$10$^{-4}$, indicating that the difference between the samples is statistically significant.
We then perform a bootstrap test analogous to that described above for the study of H$_{2}$
masses: in $\sim$87$\%$ of the iterations (8761/10000) the $\Delta$ t$_\mathrm{{dep}}$ of galaxy pairs and control galaxies are drawn
from a different distribution with a probability $\geq$99.97$\%$ (which corresponds to a difference $\geq$3$\sigma$, as previously done in the analysis of $\Delta$M$_{H_{2}}$).

However, when the two control sources from xCOLDGASS are also matched in SFR, the median $\Delta$t$_\mathrm{{dep}}$ is 0.09 dex (bottom panel of Figure 5). A KS test applied to the `depletion time offset' distributions drawn from the original samples of galaxy pairs and isolated galaxies from xCOLDGASS indicates the they are statistically indistinguishable (p-value=0.24), and this is confirmed by the bootstrap test which shows that
in none of the 10000 iterations the null hypothesis that two artificial samples are drawn from the same underlying distributions can be rejected.
% which implies that depletion times
%in our sample are consistent with those of normal galaxies with the same  
%physical `control' properties. 

Thus, our sources appear to exhibit both higher gas masses and shorter depletion times when compared to normal galaxies  compared with a mass matched sample of non-interacting galaxies drawn from xCOLDGASS. However, when matched additionally matched in SFR, these differences seem to disappear.

\begin{figure*}
\centering
\includegraphics[width=120mm,angle=0]{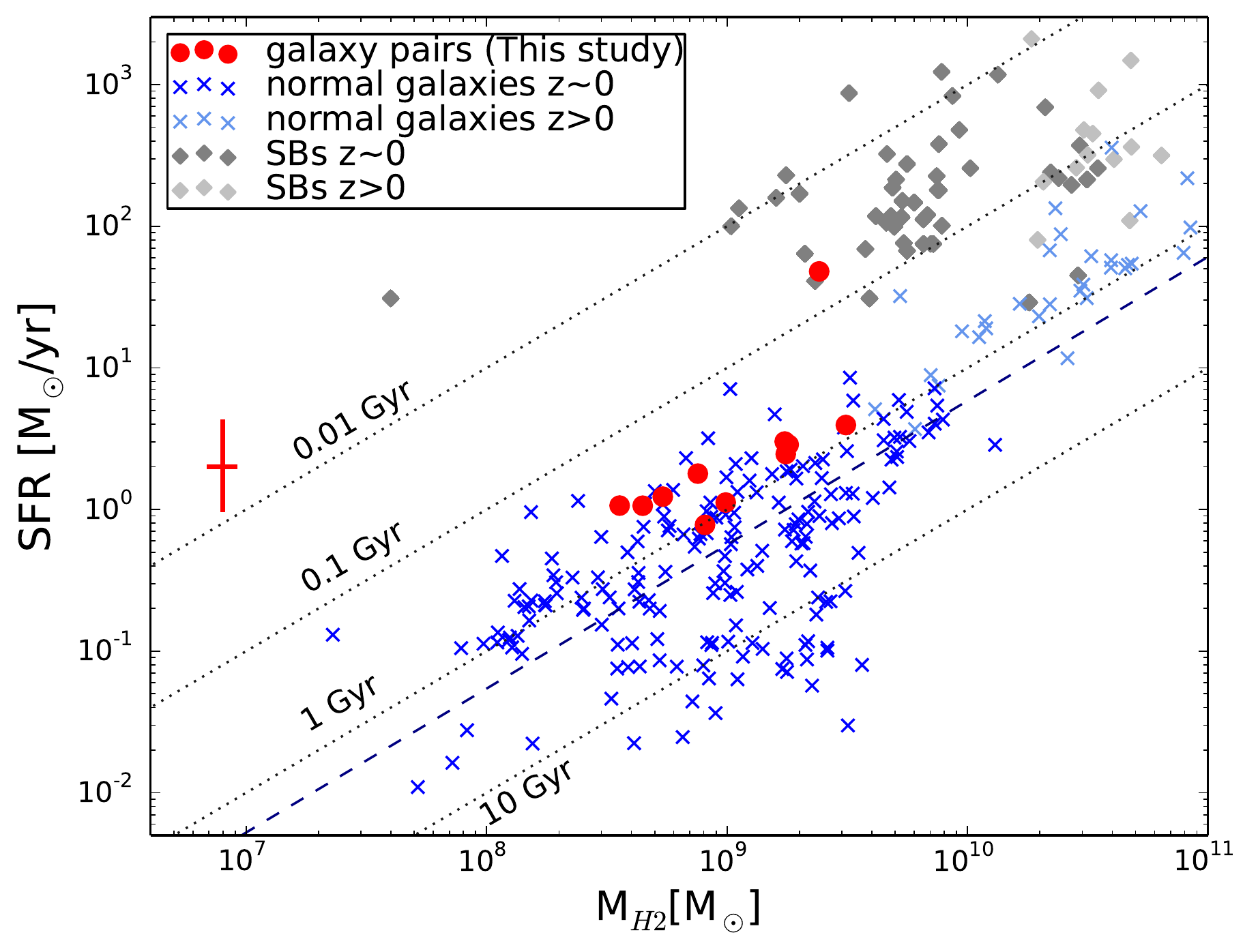}
\caption{SFR plotted as a function of the molecular gas mass M$_{\mathrm{H_{2}}}$. Our 11 galaxies in pairs are represented as red circles, while normal star-forming galaxies and starbursts are blue crosses and grey diamonds respectively. For both literature samples the dark symbols are local/low-redshift galaxies. The red cross
reproduces the average error on M$_{H_{2}}$ and SFR of our sample. For reference we show lines of constant gas depletion time. The dashed blue line is the best fit to the normal galaxies; with galaxy pairs lying above this sequence (with an average offset $\leq$1$\sigma$) as a result of their enhanced SFE.}
\end{figure*}

\section{Discussion}

The principal aim of this study is to investigate the connection between the gas properties and the star formation in galaxies at an early stage of a major merger. 
For this reason we conducted a study of the molecular gas content
and depletion time of a sample of 11 local galaxy pairs which display high sSFRs, as expected for galaxies with a close interacting companion.
This is the first time (to our knowledge) that the gas properties of local galaxy pairs have been
studied through a systematic comparison with that of normal galaxies, performing a homogeneous analysis for the merger and control samples. 
Firstly, our method ensures that the main physical quantities we analysed, e.g., M$_{\mathrm{H_{2}}}$ and SFR, are calculated consistently
between the sample of galaxy pairs and the comparison (control) sample to minimize any systematics. Secondly, we have used a physically motivated conversion factor to 
derive M$_{\mathrm{H_{2}}}$ from our CO observations. Again, this is done consistently for both the pairs and the control sample.
Finally, we have constructed a sub-sample of control galaxies by carefully matching to stellar mass, SFR, redshift and local density, so that we can robustly assess any offset between our pairs sample and the control sample.

\subsection{Comparison with previous studies}

We find that intensified star formation in galaxy pairs appears to be associated with an increase in the molecular gas content relative to the total mass by $\sim$ 0.4 dex (Figures 4 and 5). This result qualitatively confirms what has been seen previously, although the extent of the observed gas content enhancement varies amongst different studies. 
\cite{Combes94} analysed a sample of 53 IRAS-detected galaxies in binary systems at redshift z$\sim$0 and found that these sources are characterized by CO(1-0) luminosities which are, on average, one order of magnitude higher compared to local spirals, which consequently translates into higher molecular gas masses.
One uncertainty in the \cite{Combes94} study is the use of one conversion factor $\alpha_{\mathrm{CO}}$=3.68 for the entire sample of galaxy pairs - a value reminiscent of the Milky Way and which could lead to an overestimation of the H$_{2}$ masses.
Indeed, we find a variation of the $\alpha_{CO}$ values in our sample with a typical value that deviates from the Galactic one by approximately a factor of two. 
Similarly, \cite{Braine93} observed low-redshift perturbed galaxies, which may reside in a more advanced phase of a merger than galaxy pairs, and argued that they contain more gas than normal disk galaxies by a factor of $\sim$2, with most of it residing in the centre of the galaxy. The same conclusion was reached by \cite{Ueda14}, who used ALMA resolved CO maps of post-mergers to show that their gas emission is mostly centrally concentrated. A slightly different result was reached by \cite{Kaneko13a}, who mapped the CO(1--0) emission in four early- and mid- stage mergers in the local Universe. They found that molecular gas in interacting galaxies is enhanced with respect to field galaxies (f$_{\mathrm{gas}} \sim$0.2 dex higher in mergers), however, they found the concentration of molecular gas to be lower in the former.\\

One of the most popular scenarios invoked to explain the enhanced molecular gas content observed in galaxy mergers envisages
the transition of galactic neutral gas (HI) to the molecular phase (\citealt{Braine93}; \citealt{Elmergreen93}; \citealt{Kaneko13b}).
Interestingly, \cite{Ellison15} carried out a similar analysis to that presented here, and showed no evidence of lower HI content  (quantified by a HI fraction offset) in galaxies undergoing a merger, nor prior to the collision, neither in the post-merger phase. Following the method presented in Sec. 4 for gas mass offsets, we can equivalently compute gas fraction offsets for the pair galaxies relative to the xCOLDGASS control sample.
In Figure 7 we combine the gas fraction offsets that we determine for the molecular gas with the atomic gas fraction offsets from Ellison et al. (2015) in order to summarize changes in molecular and atomic gas fractions as a function of merger stage.  As we previously found for molecular gas masses (Sec 4.4) the H$_2$ gas fraction is elevated when compared to a mass matched sample, but consistent with the control when additionally matched in SFR.  Sargent et al. (in prep) have performed a similar analysis to the one presented here, but using a sample of post-merger galaxies, representing a later stage in the merger sequence than the pairs sample.  Sargent et al. (in prep) measure an enhanced molecular gas fraction of $\sim$ 0.6 dex relative to a mass and redshift matched control sample, qualitatively reproducing the enhanced molecular gas fraction found in our pairs sample.  However, whereas the H$_2$ gas fraction enhancement in the pairs is no longer significant when the elevated SFRs are taken into account, the post-merger sample studied by Sargent et al. (in prep) shows a persistent $\Delta f_{gas} \sim 0.2$ even when the elevated SFRs are matched in the control sample.

%Overall, neither of these studies found a significant change in the gas fraction content of mergers, once a rigourous control matching procedure had been followed. 
%It must be noted however, that the sample of post-mergers were not pre-selected to exhibit enhanced 
%sSFR, and that the control isolated galaxies employed in the comparison analysis were not matched in SFR.
%Regarding the study presented here, it cannot be excluded that the galaxies we observed were more gas rich before the beginning of the merger,
%and that therefore the interaction with a close companion has little or no visible effect on their molecular gas content. 

Our next finding is that enhanced SFR in pairs is accompanied by a reduction of the time necessary to deplete the gas (t$_{dep} \sim$0.6 Gyr), which is about $\sim$0.5 dex shorter than in normal galaxies, 
represented by the whole xCOLDGASS sample. %\textbf{SLE:  Giulio - you say 0.5 dex here, but delta tdep was -0.2, so this seems inconsistent.  Change 0.5 to 0.2?}.
\cite{Combes94} found that their sample of pairs exhibited 
depletion times up to $\sim$0.5 dex shorter than normal spirals. However the depletion times in Combes et al. are likely underestimated because of the choice of a disk-like CO--H$_{2}$ conversion factor for their entire sample. \cite{Saintonge12} analysed the depletion times of a sub-sample of COLDGASS sources classified as mergers based on their morphological features; this class of object has a mean depletion time of the order of 0.7 Gyr, a value which agrees with that found by our analysis. 
\cite{Goncalves14} studied a sample of 6 Lyman break analogues (LBAs), UV-selected star-forming galaxies in the local Universe. All their sources reside in galaxy pairs and their gas components constitute up to the 60$\%$ of the total galaxy mass. However,
 despite lying along the sequence of normal star-forming galaxies in the Schmidt-Kennicutt plane, these galaxies will deplete their gas in less than $\sim$1 Gyr.
 It must be noted however that
for these types of sources, the calculation of M$_{\mathrm{H_{2}}}$ can be particularly problematic, especially because of the uncertainties which affect the estimate of the CO--H$_\mathrm{{2}}$ conversion factor. Indeed, these sources have both low-metallicity and rather high SFRs, two characteristics which alter in opposite directions the derivation of $\alpha_{\mathrm{CO}}$ which can vary by a factor of $\sim$10 (\citealt{Leroy08}, \citealt{Papadopulos12}).
\cite{Casasola04} instead found galaxy mergers to have the same molecular gas depletion times of normal galaxies (t$_{\mathrm{dep}} \sim$1 Gyr). However, their sample was heterogeneous, including both post-mergers (i.e. galaxies exhibiting tidal features and disturbed structure) and galaxies in pairs. In addition,
they also estimated molecular gas mass utilising a single value of
$\alpha_{\mathrm{CO}}$ for all their sources, regardless of the merger-phase.

Importantly, we have also shown in Sections 4.4 and 4.5 that the are no differences in molecular gas content and depletion time between galaxy pairs and normal galaxies with matched physical properties (i.e. M$_{*}$, z, SFR and $\delta_{\mathrm{5}}$).
This however does not contradict previous results; instead it indicates that an increase in the gas content and a reduction of the depletion time are also observable in normal galaxies, and \textit{mergers are only one of the possible processes capable of inducing
these effects}.
For instance, stellar bar instabilities can cause an increase in the amount of dense gas and consequently drive enhanced 
star formation and a decrease of the gas consumption timescale (e.g. \citealt{Sheth05}; \citealt{Wang12}).

\subsection{The bimodality of the Schimidt-Kennicutt relation}

It has been suggested that normal star-forming galaxies (disks and their high-z counterparts, BzK galaxies) and starbursts (local ULIRGs and SubMillimetre galaxies), form two different sequences in the SFR--M$_{\mathrm{H_{2}}}$ plane (\citealt{Daddi10}, \citealt{Genzel10}), with the latter having molecular gas depletion times up to $\sim$2 orders of magnitude shorter. However, more recently \cite{Saintonge11b} showed that the population of LIRGs can bridge the gap between the two sequences. In addition, \cite{Sargent14} proposed that an apparent bimodality can arise because of poor sampling of intermediate sources, due to the fact the CO observing campaigns often favour either extreme star-forming objects or normal star-forming galaxies. 
We now wish to put our study in this context, 
and place our galaxy pairs in the integrated Schmidt-Kennicutt plane. To do this, we assemble a combination of normal galaxies and starbursts from the literature.
We start with the compilation
of \cite{Sargent14}, who selected 131 MS galaxies at redshift z $\leq$ 3. 
To these, we add the subset of the xCOLDGASS sample described in 4.3 as well as local ULIRGs from the works of \cite{Solomon97} and \cite{Combes13}. Moving to high-$z$, we include starburst galaxies from \cite{Bothwell13}, \cite{Rowlands15} and \cite{Silverman15}.
We restrict these comparison samples by selecting only those sources which have observations of transitions not higher than CO(2--1), as this avoids uncertainties related to excitation correction when estimating the luminosity of the ground state CO(1--0) transition (to the contrary, \citealt{Daddi10} rely on previously observed sources which span a wide variety of CO transitions, from 1--0 up to 9--8). For sources with CO(2--1) measurements we apply an excitation correction of 0.85 (\citealt{Daddi15}).
Our final comparison sample is made up of 277 sources in the redshift range 0.02--4, of which 216 are normal galaxies and 61 are starbursts.
We also point out that in deriving the value of M$_{\mathrm{H_{2}}}$ for each of these sources, we adopt a conversion factor $\alpha_{\mathrm{CO}}$
calculated with the same method used for our sample and described in Section 4.1. 

In Figure 8 we plot SFR as a function of H$_{\mathrm{2}}$ mass for the  composite literature sample together with our 11 galaxies in pairs.
Determining the exact parametrization of the two sequences is beyond the scope of this paper, but we can examine the position of our sample relative to the two sequences. The 11 galaxy pairs lie systematically above the general relation defined by normal star-forming galaxies, as expected from the previous analysis, which showed that our sources have shorter depletion times when compared with the entire xCOLDGASS sample
at fixed stellar mass.
However, it can be also noted that a significant number of
of `normal' galaxies exhibit similar SFRs to those of galaxy pairs 
(at fixed H$_{2}$ mass). This explains why, once that SFR is included as
an extra matching parameter, the depletion times of galaxy pairs and control galaxies become comparable. \\
The locus of our galaxy pairs in the SFR--M$_{\mathrm{H_{2}}}$ plane is in agreement with that predicted by previous theoretical studies. 
%\textbf{SLE: This seems contradictory to the lack of offset when SFR is matched?}
For instance, \cite{Renaud14} performed pc-scale hydrodynamical simulations of a galaxy merger, following the evolution in the Schmidt-Kennicutt plane of one of the 
interacting galaxies. According to their model, galaxies which are at the early stage of the merger, as in the case of our sources,  display only a modest elevation above the sequence of disk galaxies. This is due to the fact that the gravitational interaction with the approaching companion is still weak and not capable of driving a drastic increase of the gas density. Consequently, as the merger proceeds, gas inflows gradually increase the surface density in the galactic nucleus, leading the source on the starburst sequence only between the second encounter and the final coalesce phase. 

To summarize, our galaxy pairs appear to partially contribute to bridging the gap between the two sequences in the SFR--M$_{\mathrm{H_{2}}}$ plane.
Sources which reside at a more advanced stage of the merger may then begin to fill in the `gap', as predicted by high-resolution hydrodynamical simulations (\citealt{Powell13}; \citealt{Renaud14}). 

\begin{figure}
\centering
\includegraphics[width=90mm,angle=0]{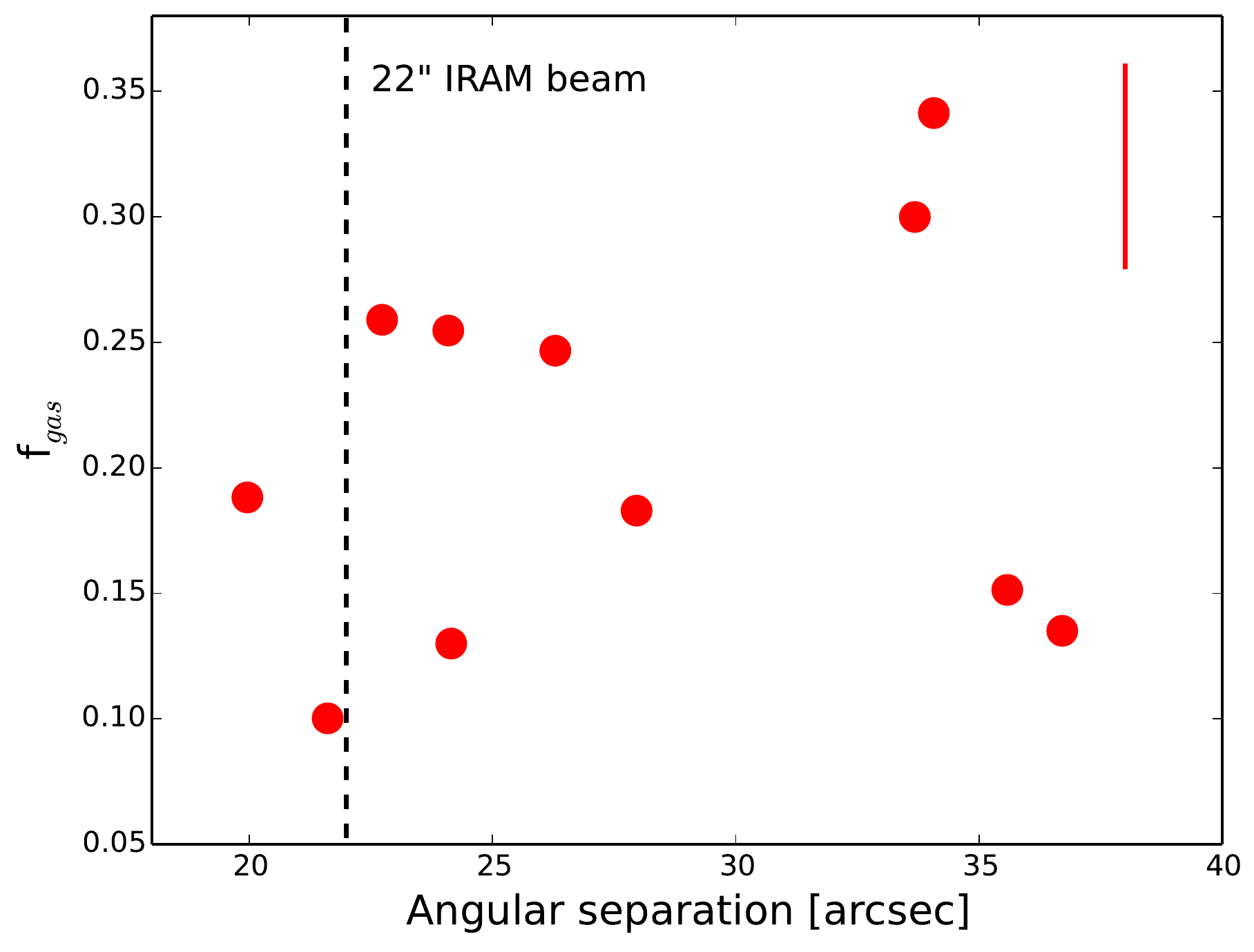}
\caption{Molecular gas fraction vs angular separation between the two component galaxies comprising each pair in our sample. The black dashed line indicates the FWHM of the IRAM 30m beam at 3 mm. The red vertical bar represents the average error on gas fraction of our sample. The majority of the sources lie above this limit, and together with the lack of correlation between the plotted quantities, suggests that our CO measurements do not suffer from contamination from gas emission of the pair companion.}
\end{figure}

\subsection{Caveats and limitations}
We now consider any possible limitations and caveats to our results.
First of all, for our CO flux measurements, we must consider the possibility of contamination from the CO emission of the companions, within the IRAM 30-m beam of 22 arcsec at 3 mm. In Figure 9 we plot the molecular gas fraction as a function of the
angular separation between the component galaxies in the pairs. 
No clear trend  between these quantities is present, suggesting no systematic contamination from the companions.
Furthermore, for the majority of our sources (9/11) the distance from the other member of the pair is well beyond the 22 arcsec (the size of the IRAM 30-m beam at 3 mm), and for the remaining two sources, the distance ($\sim$20 arcsec) is such that the potential contamination would only be marginal. 

Secondly, amongst xCOLDGASS galaxies we only select sources with CO detections, meaning that we are biased towards relatively gas-rich galaxies. However, we verified that the majority of the non-detected sources 
possess low SFR (log SFR $\leq$-0.2) and high stellar mass (log(M$_{*}$/M$_{\odot}$)$\geq$ 10), and therefore
cover a different parameter space than that of our galaxy pairs. As a consequence, their inclusion would not have an impact on our results.  Similarly, the inclusion of CO-undetected  
sources in the SFR--M$_{\mathrm{H_{2}}}$ plane (Figure 8) would alter the slope of the general relation defined by the xCOLDGASS sample, 
however our galaxy pairs would still lie well above it.

Another caveat which may affect our study is the calculation of the
 CO--H$_{\mathrm{2}}$ conversion factor. 
%Two caveats may still affect our study: the CO--H$_{2}$ conversion factor and the different SFR indicators.
To date, the most recent models in the literature which
attempt to provide a reliable value of $\alpha_{\mathrm{CO}}$ are metallicity-dependent (e.g. \citealt{Wolfire10}, \citealt{Narayanan12}, but see \citealt{Bolatto13} for a review on this topic).
Similarly, the model we employed expresses the value of $\alpha_{\mathrm{CO}}$ as a function of metallicity, SFR and stellar mass.
Importantly, the conversion factors have been estimated consistently between the sources of our sample and those of xCOLDGASS, therefore any systematics in our calculation of $\alpha_{\mathrm{CO}}$ would affect the two samples in the same way. \\
Lastly, we want to stress that our results are not applicable to the entire population of merging galaxies, but only those galaxies which, through the interaction, gain a boost in their star formation activity. Both hydrodynamical simulations (\citealt{DiMatteo08}, \citealt{Powell13}) and observations (e.g. \citealt{Saintonge12}) demonstrate that not all galaxy mergers are intrinsically associated to starbursts episodes.

\section{Conclusions}

In this paper we present the results derived from our IRAM 30-m CO(1--0) and CO(2--1) observations of 11 SDSS-selected galaxies in pairs at z$\sim$0.03. These sources represent the early stage of a major merger, and exhibit higher sSFR compared to main-sequence star-forming galaxies, as expected for galaxies undergoing an interaction. We study the molecular gas (H$_{2}$) properties of these interacting systems through a comparison with a carefully-selected control sample from xCOLDGASS (\citealt{Saintonge17}).
We find that: 
\begin{itemize}

\item The molecular gas represents, on average, at least $\sim$16$\%$ of the total stellar mass of our sources. 
 Our sample of galaxy pairs exhibits molecular gas fractions f$_{\mathrm{gas}}$=M$_{\mathrm{H_{2}}}$/M$_{*}$ which are  $\sim$0.4 dex higher than those of normal star-forming galaxies as sampled by xCOLDGASS galaxies.  The enhanced (s)SFR seen in our galaxy pairs is therefore most likely driven by both a larger gas reservoir available for fuelling star formation, and by a higher efficiency in converting this gas into stars.

\item The average molecular gas consumption timescale of our galaxy pairs is $\sim$0.6 Gyr. Compared with a mass matched control sample from xCOLDGASS we find depletion times to be 0.2 dex shorter in pairs than in non-interacting galaxies. This decrease in the molecular gas depletion time reflects an enhancement of the efficiency in converting gas into stars; this is likely due to a faster transition of the molecular gas to its denser phase, driven by the gravitational interaction with a close companion.   

\item  If we additionally match the control sample in SFR, the molecular gas fractions and depletion times are consistent between the pairs and non-interacting galaxies.  This suggests that even in normal galaxies, internal mechanisms (e.g. bar instabilities) can drive the same effect produced by galaxy interactions, such as enhancement of the molecular gas content, increase in SFR and reduction of the molecular gas depletion time. 
\end{itemize}

The results obtained in this paper can be used as a starting point for expanding the study of the gas properties in local galaxy mergers, and thus
improve our view of the ISM conditions in these `intermediate' class of star-forming galaxies. For this scope additional observations are needed.

First of all, the companions of the 11 galaxies analysed in this study can be targeted in follow-up CO(1-0) and CO(2-1) observations,
taking advantage of the successful observing strategy adopted here.
In this way, it will be possible to study the effect of the merger on the molecular gas content of both the galaxies which make up a pair, gaining a complete view of this process. 
Secondly, observations of emission lines from higher CO transitions (e.g. CO(3-2) and CO(4-3)) of the 11 galaxy pairs of this sample would provide an insight into the physical conditions of the gas.
In fact, through the study of the CO spectral line energy distribution (SLED), and a comparison with that of normal galaxies and starbursts (both at low- and high-redshift, e.g. Daddi et al. 2015) some crucial information such as gas temperature and surface densities can be inferred (e.g. \citealt{Lagos12}).
Lastly, it has been suggested that the apparent bimodality in the integrated Schmidt-Kennicutt relation disappears once only the denser component of the molecular gas is considered (\citealt{Gao07} and reference therein). An expansion of our study to induce a probe of the dense phase (3$\times$10$^{4}$ cm$^{-2}$) of the molecular gas (e.g. HCN) could shed some light on the apparently different global star formation laws that govern different types of 
sources.

\section*{Acknowledgments}

GV acknowledges the University of Hertfordshire for a PhD studentship. KEKC acknowledges support from the UK Science and Technology Facilities Council [STFC; grant numbers ST/M001008/1, ST/J001333/1]. 
JMS acknowledges support from the UK STFC Council [grant number ST/L000652/1]. 

This work is based on observations carried out with the IRAM 30-m telescope. IRAM is supported by INSU/CNRS (France), MPG (Germany) and IGN (Spain).
The authors also acknowledge the IRAM staff for help provided during the observations and for data reduction. 
Funding for the SDSS and SDSS-II has been provided by the Alfred P. Sloan Foundation, the Participating Institutions, the National Science Foundation, the U.S. Department of Energy, the National Aeronautics and Space Administration, the Japanese Monbukagakusho, the Max Planck Society, and the Higher Education Funding Council for England. The SDSS Web Site is http://www.sdss.org/.
The SDSS is managed by the Astrophysical Research Consortium for the Participating Institutions. The Participating Institutions are the American Museum of Natural History, Astrophysical Institute Potsdam, University of Basel, University of Cambridge, Case Western Reserve University, University of Chicago, Drexel University, Fermilab, the Institute for Advanced Study, the Japan Participation Group, Johns Hopkins University, the Joint Institute for Nuclear Astrophysics, the Kavli Institute for Particle Astrophysics and Cosmology, the Korean Scientist Group, the Chinese Academy of Sciences (LAMOST), Los Alamos National Laboratory, the Max-Planck-Institute for Astronomy (MPIA), the Max-Planck-Institute for Astrophysics (MPA), New Mexico State University, Ohio State University, University of Pittsburgh, University of Portsmouth, Princeton University, the United States Naval Observatory, and the University of Washington.

\setlength{\bibhang}{2.0em}

\end{document}